\documentclass[floatfix,aps,preprintnumbers,amsmath,amssymb,nofootinbib,superscriptaddress,showkeys,showpacs]{revtex4}

\usepackage{graphicx,color}
\usepackage{amsmath,amssymb,bm}
\usepackage{dcolumn}

\newcommand{\bea}{\begin{eqnarray}}
\newcommand{\eea}{\end{eqnarray}}
\newcommand{\be}{\begin{equation}}
\newcommand{\ee}{\end{equation}}
\newcommand{\np}{{\bf p}}

\newcommand{\nh}{{\bf h}}

\newcommand{\nk}{{\bf k}}

\newcommand{\nq}{{\bf q}}

\newcommand{\Qbar}{\not{\!Q}}

\newcommand{\kbar}{\not{\!k}}
\newcommand{\Pbar}{\not{\!P}}
\newcommand{\tauvec}{\mbox{\boldmath $\tau$}}

\newcommand{\Ivec}{\mbox{\boldmath $I$}}

\def\XXint#1#2#3{{\setbox0=\hbox{$#1{#2#3}{\int}$}
     \vcenter{\hbox{$#2#3$}}\kern-.5\wd0}}

\def\1{\'{\i}}

\begin{document}

\title{
Semiempirical formula for electroweak response functions in the two-nucleon emission channel in neutrino-nucleus scattering 
}

\author{V.L. Martinez-Consentino}\email{victormc@ugr.es} 
  \affiliation{Departamento de
  F\'{\i}sica At\'omica, Molecular y Nuclear \\ and Instituto Carlos I
  de F{\'\i}sica Te\'orica y Computacional \\ Universidad de Granada,
  E-18071 Granada, Spain.}

\author{J.E. Amaro}\email{amaro@ugr.es} \affiliation{Departamento de
  F\'{\i}sica At\'omica, Molecular y Nuclear \\ and Instituto Carlos I
  de F{\'\i}sica Te\'orica y Computacional \\ Universidad de Granada,
  E-18071 Granada, Spain.}

\author{I. Ruiz Simo}\email{ruizsig@ugr.es} \affiliation{Departamento de
  F\'{\i}sica At\'omica, Molecular y Nuclear \\ and Instituto Carlos I
  de F{\'\i}sica Te\'orica y Computacional \\ Universidad de Granada,
  E-18071 Granada, Spain.}

\date{\today}

\begin{abstract}
\rule{0ex}{3ex} 

A semi-empirical formula for the inclusive electroweak response functions in the
two-nucleon emission channel is proposed.  The method consists in
expanding each one of the vector-vector, axial-axial and vector-axial
responses as sums of six sub-responses.  These correspond to
separating the meson-exchange currents as the sum of three currents of
similar structure, and expanding the hadronic tensor as the sum of
the separate contributions from each current plus the interferences
between them. For each sub-response we factorize the coupling
constants, the electroweak form factors, the phase space and the delta
propagator, for the delta forward current.  The remaining spin-isospin
contributions are encoded in coefficients for each value of the momentum
transfer, $q$.  The coefficients are fitted to the exact results in the
relativistic mean field model of nuclear matter, for each value of
$q$. The dependence on the energy transfer, $\omega$ is well described
by the semi-empirical formula. The $q$-dependency of the 
coefficients of the sub-responses can
be parameterized or can be interpolated from the provided tables. The
description of the five theoretical responses is quite good. The
parameters of the formula, the Fermi momentum, number of particles,
relativistic effective mass, vector energy, the electroweak form
factors, and the coupling constants, can be modified easily. 
 This semi-empirical formula can be applied to the
cross-section of neutrinos, antineutrinos and electrons.

\end{abstract}



\maketitle

\section{Introduction}

In the last years the study of quasi-elastic neutrino scattering by
nuclei has received increasing interest from the theoretical and
experimental point of view. This is due to recent neutrino accelerator
experiments aimed at measuring  neutrino-nucleus cross sections.
The ultimate goal is to measure the neutrino oscillation parameters accurately,
something that requires knowing with precision the cross-sections of
neutrino scattering with the nuclei of the detectors. Extensive reviews
 describing the state of the art of the subject
 namely, recent advances and open
challenges in the field of neutrino-nucleus scattering,  can be found in Refs.
\cite{Gal11,Mor12,For12,Alv14,Mos16,Ank17,Ben17,Kat18,Ama20}.

Experimental measurements of the inclusive charge-changing (CC) quasi-elastic
cross-section $(\nu_\mu,\mu^-)$ and $(\overline{\nu}_\mu,\mu^+)$, have
been carried out in several laboratories, each of which is
characterized by a neutrino flux or distribution in a range of
energies with different widths around 1 GeV
\cite{Nomad09,Agu10,Agu13,Fio13,Abe13,Abe16,Abe18}.  Comparison of
these data with the different nuclear models of the reaction revealed
the discrepancies between the different approaches, as well as
discrepancies with the experimental data
\cite{Mar09,Nie11,Gal16,Meg16,Meg14,Ank15,Gra13,Pan16,Mar16}.  In
particular, the importance of including the contributions of
multinucleon emission was highlighted.  The calculation of 2p2h
contribution varies in different models, but all agree on a
contribution that can be around 20\% of the cross section with
one-nucleon emission only.
                          
Neutrino CC cross sections are averages for different incident
energies. Therefore, the differences between models are also
differences between averages ---although theoretically the double
differential cross section is calculated for fixed neutrino
energies. To compare with data for fixed energy the only possibility
is to use the electron $(e,e')$ experiments. This allows the neutrino
models to be calibrated, taking into account that the electromagnetic
current is related to the vector part of the weak current. It is only
the axial current that is not fixed by the electrons. Various
approaches to the study of quasi elastic lepton scattering have been
applied. For finite nuclei  quantum Monte Carlo methods \cite{Lov16},
continuum shell model with RPA correlations \cite{Pan15}, or spectral
function methods \cite{Ank15b,Roc16} can be mentioned.  For high or
intermediate energies relativistic corrections are important, yet
corrections to kinetic energy can be partially included. But the
nuclear interaction at the relativistic level requires special care
\cite{Ama02,Ama05}. The relativistic mean field (RMF) is the simplest
model available that includes the interaction in a fully relativistic
way in the form of scalar and vector mean potentials
\cite{Cab07,Udi99}.  Other models use fits to generate
e.g. parametrizations of the transverse enhancement of the transverse
response \cite{Bod14,Gal16}.  Most of the approaches require
approximations that sometimes violate fundamental current conservation
requirements. The influence of other factors such as short-range
correlations, final-state interaction, exchange currents, pion
emission, and resonance excitation contribute to the differences
between the various models and to the difficulty of accurately
reproducing electron data in the full kinematical range \cite{Sob17}.

In this paper we focus on the inclusive two-nucleon emission (2p2h)
channel for neutrino and electron scattering. Due to the nuclear force
between correlated nucleons, a high-energy neutrino can induce
emission of a pair of nucleons. In the case of $(\nu_\mu,\mu^-)$
reaction, PP and PN pairs can be ejected, while NN and PN pairs can be
ejected in the $(\overline{\nu}_\mu,\mu^+)$ reaction.  A large part of the
2p2h cross section is produced by interaction with nuclear
meson-exchange currents (MEC), where the exchanged bosons $W^\pm$ are
absorbed by mesons exchanged between two nucleons.  The most important
part of the 
interaction occurs when the boson absorption by the nucleon produces a
virtual $\Delta$ that decays communicating its energy to a pair of
nucleons.

Several phenomenological models have been proposed to calculate
the 2p2h channel in neutrino interactions.  Each of these models, based on
the Fermi gas, takes its own approximations and in general they are
numerically expensive.  A first calculation revealed the importance of
2p2h to reproduce the data of neutrino scattering \cite{Mar09,Mar10},
and similar effects were found in a relativistic calculation
\cite{Ama11,Ama12} and also in a local Fermi gas with random-phase
approximation (RPA) and other many-body corrections
\cite{Nie11,Nie12,Nie13}. These models require integration in many
dimensions and intensive numerical computation of traces of two-body
operators.  Although several approaches have been proposed
\cite{Sim17b,Sim18b} that reduce the calculation, in the absence of
analytical formulas for the 2p2h responses it is necessary to
elaborate numerical tables or parametrizations of theoretical
calculations for their implementation in Monte-Carlo simulators
\cite{Sob20,Dol20}.

In ref. \cite{Lal12} the impact of 2p2h was investigated in a simple
model using an ansatz physically motivated by the phase space function
of two particles in a Fermi gas. The matrix elements of the MECs
averaged over the neutrino flux were approximated by a constant that
was fitted to the neutrino data. In this work we explore extensions of
this simple model by adding other physical dependencies, motivated
by the theoretical structure of the MEC operators, which contain form
factors, coupling constants and propagators of the $\Delta$, as
physical dependencies that must be present in the MEC responses. The
rest of the more complex dependencies associated with elements of the
spin-isospin matrix of Dirac operators are synthesized in constants
for each value of the transferred momentum. These constants are obtained
by a fit of a theoretical model, but could also be fitted to data if
there were a sufficient number of them.  The resulting semi-empirical (SE)
formula for the MEC responses allows to compute the inclusive 2p2h
responses analytically for fixed $q$, having the parameters tabulated
as a function of $q$, and to interpolate for intermediate values.

The need of an analytical formula for the inclusive 2p2h responses is based on
the difficulty of calculating the integral in seven dimensions for the
2p2h responses, which then has to be integrated with the neutrino
flux. With a suitable parameterization the calculation time is
reduced. An alternative parameterization was carried out in the
relativistic Fermi gas (RFG) model of ref. \cite{Meg15} for the vector
responses, containing gaussians and polynomials. A similar
parametrization was made in ref. \cite{Meg16} for the axial
contributions. In this paper we follow a different approach, using the
relativistic mean field (RMF) model of nuclear matter, and a
semi-empirical parameterization of the 2p2h responses that has the
advantage that the dependence on the momentum, $q$, and energy
transfer, $\omega$, has a theoretical basis. In analogy with the
Bethe-Weizsacker semi-empirical mass formula of nuclear binding energy
\cite{Wei35}, where the dependence on A and Z also has a theoretical
basis. In particular, we extract the dependence of the responses on
the number of particles, the Fermi momentum, and the relativistic effective
mass. Furthermore, the dependence is exact on the form factors and the
electroweak coupling constants. The formula will contain constants
that we will adjust to the exact results.  The frozen and modified
convolution (MCA) approximations have been proposed
\cite{Sim17b},\cite{Sim18b} that reduce the integrals to one dimension
and five, respectively, to which an integral over the neutrino flux
would have to be added. However, the semi-empirical formula has the
advantage that it is analytical and gives better results than the
frozen one, since the parameters have been adjusted to reproduce the
exact responses.

Note that the developments in this paper are applicable to the
calculation of inclusive cross sections only.
The MEC model that we use to fit the SE-MEC responses is based on
Feynman diagrams for pion production on the nucleon. In our case these
diagrams will be considered in a relativistic mean field model of
nuclear matter. In this model the nucleons are relativistic and
interact with vector and scalar potentials, the effect of which is to
give them an effective mass and a repulsive vector energy.  The idea
is to combine this SE-MEC formula with the super scaling approach with
relativistic effective mass (SuSAM*) were a new scaling function
$f^*(\psi^*)$ including dynamical relativistic effects has been
proposed \cite{Ama15,Ama17,Mar17} through the introduction of an
effective mass for the nucleon.  The SuSAM* model describes a large
amount of the electron data within a phenomenological quasielastic
band, and it was extended to the neutrino and antineutrino sector
\cite{Rui18}, SuSAM* was first developed from the set of $^{12}$C data
\cite{Ama15,Ama17} and later applied to other nuclei in
\cite{Mar17,Ama18}. Recently the super scaling function has been
refitted by subtraction of the MEC 2p2h cross section before
performing the scaling analysis, in order to avoid double counting
when adding the MEC responses \cite{Mar21}.

The scheme of the paper is as follows.  In Sect. II we present the
formalism for neutrino scattering, the inclusive 2p2h hadronic tensor,
the model of MEC, the analytical approximation of the phase space
integral in frozen nucleon approximation, and the averaged $\Delta$
propagator.  In Sect. III we write the semi-empirical formula for the
MEC responses.  In sect IV we present the results of the fit and
compare with the exact results.  In Sect. V we draw our conclusions.

\section{Formalism of CC neutrino scattering}
\label{sec_form}

\subsection{Neutrino cross section}

In this section we summarize the formalism for charge-changing 
neutrino scattering. The case of
electron scattering can be easily inferred from this by considering
only the longitudinal and transverse response functions.  Thus
we consider charged-current inclusive quasielastic (CCQE) reactions in
nuclei induced by neutrinos and antineutrinos, focusing on the
  $(\nu_\mu,\mu^-)$ and  $(\overline\nu_\mu,\mu^+)$ 
  cross sections.  The relativistic
energies of the incident (anti)neutrino and detected muon are
$\epsilon=E_\nu$, and $\epsilon'=m_\mu+T_\mu$, respectively. Their
momenta are $\nk$ and $\nk'$.  The four-momentum transfer is
$k^\mu-k'{}^\mu=Q^\mu=(\omega,\nq)$, with $Q^2=\omega^2-q^2 < 0$.  The
lepton scattering angle, $\theta$, is the angle between $\nk$ and
$\nk'$.  The double-differential cross section can be written as
\begin{eqnarray}
\frac{d^2\sigma}{dT_\mu d\cos\theta}
=
\frac{G^2\cos^2\theta_c}{4\pi}
\frac{k'}{\epsilon}v_0  
\left[V_{CC} R_{CC}+ 
2{V}_{CL} R_{CL}+
{V}_{LL} R_{LL}+
{V}_{T} R_{T}
\pm
2{V}_{T'} R_{T'}
\right] \, .
\label{cross}
\end{eqnarray}
Inside the brackets in Eq. (\ref{cross}) 
there is a linear combination of the five nuclear
response functions, where (+) is for neutrinos and $(-)$ is for
antineutrinos.  
Here $G=1.166\times 10^{-11}\quad\rm MeV^{-2}$ is
the Fermi constant, $\theta_c$ is the Cabibbo angle,
$\cos\theta_c=0.975$, and the kinematical factor $v_0=
(\epsilon+\epsilon')^2-q^2$.
The $V_K$ coefficients depend only on the 
lepton kinematics 
and do not depend on the details of the nuclear target:
\begin{eqnarray}
{V}_{CC}
&=&
1+\delta^2\frac{Q^2}{v_0}
\label{vcc}\\
{V}_{CL}
&=&
\frac{\omega}{q}-\frac{\delta^2}{\rho'}
\frac{Q^2}{v_0}
\\
{V}_{LL}
&=&
\frac{\omega^2}{q^2}-
\left(1+\frac{2\omega}{q\rho'}+\rho\delta^2\right)\delta^2
\frac{Q^2}{v_0}
\\
{V}_{T}
&=&
-\frac{Q^2}{v_0}
+\frac{\rho}{2}+
\frac{\delta^2}{\rho'}
\left(\frac{\omega}{q}+\frac12\rho\rho'\delta^2\right)
\frac{Q^2}{v_0}
\\
{V}_{T'}
&=&
-\frac{1}{\rho'}
\left(1-\frac{\omega\rho'}{q}\delta^2\right)
\frac{Q^2}{v_0},
\label{vtp}
\end{eqnarray}
where  we have defined the dimensionless factors
$\delta = m_\mu/\sqrt{|Q^2|}$, proportional to the muon mass $m_\mu$, 
$\rho = |Q^2|/q^2$, and $\rho' = q/(\epsilon+\epsilon')$.

The response functions, $R^K(q,\omega)$,
are defined as suitable combinations of the hadronic tensor, $W^{\mu\nu}$, in a
reference frame where the $z$ axis ($\mu = 3$) points along the momentum transfer
$\nq$, and the $x$ axis ($\mu = 1$) is defined as the transverse (to $\nq$) component of
the (anti)neutrino momentum $\nk$ lying in the lepton scattering plane; the $y$ 
axis ($\mu = 2$) is then normal to the lepton scattering plane. The usual components are then
\begin{eqnarray}
R^{CC} &=&  W^{00} \label{rcc} \\
R^{CL} &=& -\frac12\left(W^{03}+ W^{30}\right) \\
R^{LL} &=& W^{33}  \\
R^{T} &=& W^{11}+ W^{22} \\
R^{T'} &=& -\frac{i}{2}\left(W^{12}- W^{21}\right)\,. \label{rtprima}
\end{eqnarray}
The inclusive hadronic tensor is constructed from the matrix elements
of the current operator $J^\mu(Q)$ between the initial and final
hadronic states, summing over all the possible final nuclear states
with excitation energy $\omega=E_f-E_i$, and averaging over the
initial spin components.
\begin{equation}
W^{\mu\nu}
= 
\sum_f \overline{\sum_i} 
\langle f | J^\mu(Q) |i \rangle^*
\langle f | J^\nu(Q) |i \rangle
\delta(E_i+\omega-E_f) .
\label{hadronic-tensor}
\end{equation}

In the case of electron scattering the cross section is
\begin{equation}
\frac{d\sigma}{d\Omega d\epsilon'}
= \sigma_{\rm Mott}
(v_L R_{em}^L +  v_T  R_{em}^T).
\end{equation}
where $\sigma_{\rm Mott}$ is the Mott cross section, 
 $v_L$ and $v_T$ are  kinematic factors
\begin{eqnarray}
v_L &=& 
\frac{Q^4}{q^4} \\
v_T &=&  
\tan^2\frac{\theta}{2}-\frac{Q^2}{2q^2}.
\end{eqnarray}
The electromagnetic longitudinal and transverse response functions, 
 $R_{em}^L(q,\omega)$ and $R_{em}^T(q,\omega)$, 
are
\begin{eqnarray}
R_{em}^L &=& W^{00}_{em}\\
R_{em}^T &=& W^{11}_{em}+W^{22}_{em}
\end{eqnarray}

Attending to the kind of final states, the hadronic tensors can be
written as sum of one-particle (1p1h) plus two-particles
(2p2h),\ldots, emission channels
\begin{equation} 
W^{\mu\nu}
= 
W^{\mu\nu}_{\rm 1p1h}
+ W^{\mu\nu}_{\rm 2p2h} + \cdots
\end{equation}
In this work we are interested in the 2p2h channel only. We studied
the 1p1h responses and cross section for CC neutrino and antineutrino
scattering in ref. \cite{Rui18} and for electron scattering in
\cite{Mar17,Ama18,Mar21}.  Here we extend these works to deepen the
study of the interaction of neutrinos with nuclei at intermediate
energies and the role of the 2p2h electroweak response functions, as
described in the next subsection.

\subsection{2p2h Hadronic tensor in the RMF}

In this paper we consider the 2p2h responses within the relativistic mean
field (RMF) model of nuclear matter \cite{Ros80,Ser86,Weh93,Bar98}. 
In this model the nucleons are 
interacting with a relativistic field containing scalar and vector potentials.
The single particle wave functions are plane waves with momentum $\nh$, 
and with on-shell energy 
\begin{equation} \label{onshell}
E= \sqrt{(m_N^*)^2+h^2} 
\end{equation}
where $m_N^*$ is the relativistic effective mass of the nucleon defined by
\begin{equation}
m_N^*=m_N-g_s\phi_0 = M^*m_N
\end{equation}
where $m_N$ is the bare nucleon mass, $g_s\phi_0$ is the scalar
potential energy of the RMF \cite{Ser86}, and $M^*=0.8$ for $^{12}$C,
the nucleus considered in our results.  Additionally the nucleon
acquires a positive energy due to the repulsion by the relativistic
vector potential, $E_v=g_vV_0$.
Thus the total nucleon energy is 
\begin{equation}  \label{energyrmf}
E_{RMF}=E+E_v.
\end{equation}
In this work we use values of $k_F$, $M^*$ and $E_v$ that are obtained
phenomenologically from the data of $(e, e ')$  \cite{Mar21}. In particular we
use the value $E_v=141$ MeV  for $^{12}$C.
 In the semi-empirical formulas in the next
section these values could easily be modified if desired.
For many observables, that only depend on the energy differences
between a final particle and an initial one, the vector energy cancels
out and does not affect the results.
 
The nuclear states in the RMF are Slater determinants constructed with
plane waves obtained by solving the free Dirac equation with effective
mass $m_N^*$.  Note that we use the same effective mass for particles
and holes. In this approximation the model is the simplest possible
that implements the mean field, so the results for very large momenta
should be taken with caution.

 All states with momentum $h<k_F$ are occupied in the
ground state, with $k_F$ the Fermi momentum.  The 2p2h excitations are
obtained by raising two particles above the Fermi level, with momenta
$p'_1$ and $p'_2>k_F$, leaving two holes with momenta $h_1$ and
$h_2<k_F$.  In this work we focus on the 2p2h part of the hadronic
tensor.  It is generated by the interaction with a two-body current
operator, whose matrix elements are given by
\begin{eqnarray}
\langle f | J^\mu(Q) | i \rangle =
\frac{(2\pi)^3}{V^2}\delta(\np'_1+\np'_2-\nq-\nh_1-\nh_2)
\frac{(m^*_N)^2}{\sqrt{E'_1E'_2E_1E_2}}
j^{\mu}(\np'_1,\np'_2,\nh_1,\nh_2),
\end{eqnarray}
where $V$ is the volume of the system and the on-shell energies and 
momenta of the
particles and holes are $(E'_i,\np'_i)$, and $(E_i,\nh_i)$,
for $i=1,2$.  Note that three-momentum is conserved, and that the
relativistic factors $(m^*_N/E)^{1/2}$ contain the effective
mass and the on-shell energies. The
spin-isospin  two-body current matrix elements
$j^{\mu}(\np'_1,\np'_2,\nh_1,\nh_2)$ are defined in the next
subsection.

Inserting this expression into the hadronic tensor, Eq.
(\ref{hadronic-tensor}),
and taking the limit
$V\rightarrow \infty$ (infinite nuclear matter) 
we transform the sums into integrals, to obtain
the 2p2h hadronic tensor for the RMF theory of nuclear matter
 \begin{eqnarray}
W^{\mu\nu}_{\rm 2p2h}
&& 
=\frac{V}{(2\pi)^9}\int
d^3p'_1
d^3p'_2
d^3h_1
d^3h_2
\frac{(m^*_N)^4}{E_1E_2E'_1E'_2}
 w^{\mu\nu}(\np'_1,\np'_2,\nh_1,\nh_2)\;
\nonumber\\
&&
\times 
\Theta(p'_1,h_1)\Theta(p'_2,h_2)
\delta(E'_1+E'_2-E_1-E_2-\omega)
\delta(\np'_1+\np'_2-\nq-\nh_1-\nh_2) 
\label{hadronic12}
\end{eqnarray}
where $V/(2\pi)^3 = Z/(\frac8 3 \pi k_F^3)$ for symmetric nuclear matter.
The Pauli blocking function $\Theta$ is defined as  
the product of step-functions for the initial and final momentum 
\begin{equation}
\kern -8mm
\Theta(p'_i,h_i) \equiv
\theta(p'_i-k_F)
\theta(k_F-h_i) .
\end{equation}

Finally the function $w^{\mu\nu}(\np'_1,\np'_2,\nh_1,\nh_2)$ represents the
hadron tensor for a single 2p2h transition,
summed up over spin and isospin,
\begin{eqnarray}
\kern -8mm
w^{\mu\nu}(\np'_1,\np'_2,\nh_1,\nh_2) &=& \frac{1}{4}
\sum_{s_1s_2s'_1s'_2}
\sum_{t_1t_2t'_1t'_2}
j^{\mu}(1',2',1,2)^*_A
j^{\nu}(1',2',1,2)_A \, . 
\label{elementary}
\end{eqnarray}
The two-body current
matrix elements is antisymetrized  
\begin{equation} \label{anti}
j^{\mu}(1',2',1,2)_A
\equiv j^{\mu}(1',2',1,2)-
j^{\mu}(1',2',2,1) \,.
\end{equation}
 The factor $1/4$ in Eq.~(\ref{elementary}) accounts for the
 anti-symmetry of the two-body wave function with respect
to  exchange of momenta, spin and isospin quantum numbers.
 
In this work we use Eq. (\ref{hadronic12}) to compute the five
2p2h inclusive response functions $R_{CC}, R_{CL}, R_{LL}, R_T, R_{T'}$, 
for neutrino and antineutrino scattering. 
Integrating over $\np'_2$ using
the momentum delta-function, these responses are given by
\begin{eqnarray}
R^{K}_{\rm 2p2h}
=
\frac{V}{(2\pi)^9}\int
d^3p'_1
d^3h_1
d^3h_2
\frac{(m_N^*)^4}{E_1E_2E'_1E'_2}
\Theta(p'_1,h_1)\Theta(p'_2,h_2)
r^{K}(\np'_1,\np'_2,\nh_1,\nh_2)\;
\delta(E'_1+E'_2-E_1-E_2-\omega) ,
\label{responses}
\end{eqnarray}
where $\bf p'_2= h_1+h_2+q-p'_1$.  The five elementary response
functions for a 2p2h excitation $r^K$ are defined in Eqs.
(\ref{rcc}--\ref{rtprima}), in terms of the elementary hadronic tensor
$w^{\mu\nu}$, Eq. (\ref{elementary}), for $K=CC,CL,LL,T,T'$.  Due to
axial symmetry around the momentum transfer $\nq$ (the $z$ axis), we
can fix the azimuthal angle of the first particle, $\phi'_1=0$ and
multiply by $2\pi$.  Finally the energy delta-function allows to
integrate analytically over $p'_1$, reducing Eq.~(\ref{responses}) to
seven dimensions integral, that we compute numerically \cite{Sim14}.

\subsection{Electroweak meson-exchange currents}

\begin{figure}[t]
\centering
\includegraphics[width=10cm,bb=110 310 500 690]{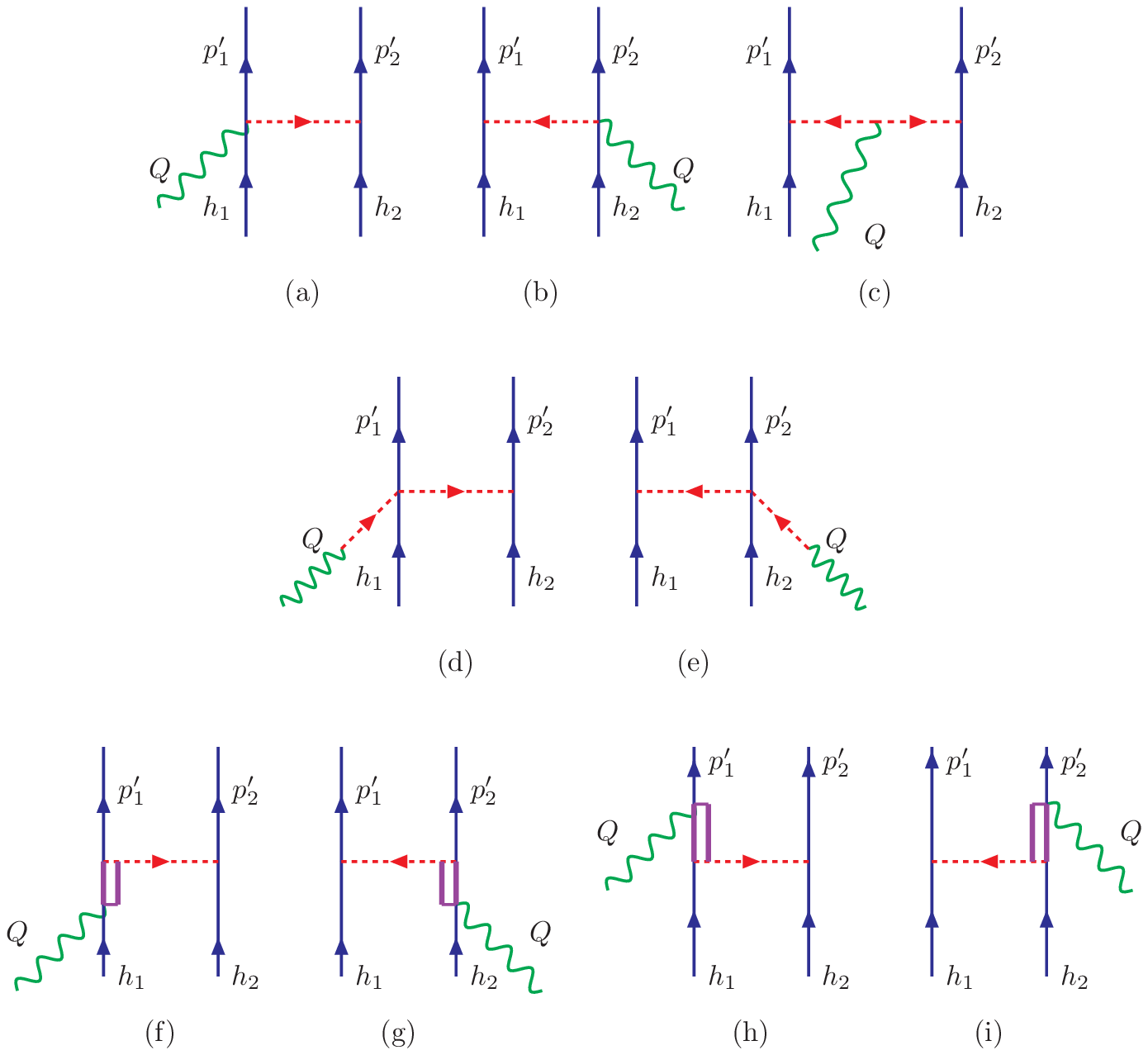}
\caption{Feynman diagrams for the electroweak MEC model used in 
this work.}\label{fig_feynman}
\end{figure}

In this work we use the electroweak MEC model described by the nine
Feynman diagrams depicted in Fig.~\ref{fig_feynman}.  The two-body
current matrix elements $j^{\mu}(1',2',1,2)$ corresponding to this
model enter in the calculation of the elementary 2p2h hadronic tensor,
Eq. (\ref{elementary}).
The different  contributions have been taken 
from the pion weak production model of ref. \cite{Hernandez:2007qq}.

To obtain the MEC we have started from the pion production currents
given in Eq (51) of ref. \cite{Hernandez:2007qq}.  and we have applied
the Goldberger-Treiman relation $f_{\pi NN}/m_\pi= g_A/(2f_\pi)$, with
$f_\pi=93$ MeV, to factorize a common coupling constant $f_{\pi NN} /
m_\pi$ in all vertices $\pi NN$. We have not included the nucleon-pole
diagrams where the $W^{\pm}$-boson is coupled to the nucleon current,
since these diagrams are not considered part of the MEC and are not
taken into account here.  
Some previous calculations of
the correlations in the 2p2h channel, of pionic type \cite{Ama10} or of 
Jastrow type \cite{Van16}, indicate that their effect when added to
the one-body response is to produce a tail to the right of the QE peak
of the order of 10\% of the total height. 

The MEC is the sum of four two-body currents: 
seagull (diagrams a,b), pion in flight (c), pion-pole (d,e) and
$\Delta(1232)$ excitation forward (f,g), and backward (h,i).
 \begin{eqnarray}
 j^\mu_{\rm sea}
&=&
  \left[I_V^{\pm}\right]_{1'2',12}
\frac{f_{\pi NN}   ^2}{m^2_\pi}
V_{\pi NN}^{s'_1s_1}(\np'_1,\nh_1) 
F_{\pi NN}(k_1^2)
 \bar{u}_{s^\prime_2}(\np^\prime_2)
 \left[ F^V_1(Q^2)\gamma_5 \gamma^\mu
 + \frac{F_\rho\left(k_{2}^2\right)}{g_A}\,\gamma^\mu
   \right] u_{s_2}(\nh_2)
   +
   (1\leftrightarrow2) \,
\label{seacur}
\\
 j^\mu_{\pi}&=& \left[I_V^{\pm}\right]_{1'2',12}
 \frac{f_{\pi NN}^2}{m^2_\pi}
 F^V_1(Q^2)
V_{\pi NN}^{s'_1s_1}(\np'_1,\nh_1) 
V_{\pi NN}^{s'_2s_2}(\np'_2,\nh_2) 
\left(k^\mu_{1}-k^\mu_{2}\right)
\label{picur}
\\
j^\mu_{\rm pole}
&=&
\left[I_V^{\pm}\right]_{1'2',12}
\frac{f_{\pi NN}^2}{m^2_\pi}\,
\frac{F_\rho\left(k_{1}^2\right)}{g_A}
F_{\pi NN}(k_2^2)
\frac{
Q^\mu
\bar{u}_{s^\prime_1}(\np^\prime_1)\Qbar u_{s_1}(\nh_1)
 }{Q^2-m^2_\pi}
V_{\pi NN}^{s'_2s_2}(\np'_2,\nh_2) 
+(1\leftrightarrow2)
\label{polecur}
\\
j^\mu_{\Delta F}
&=&
\left[U_{\rm F}^{\pm}\right]_{1'2',12}
\frac{f^* f_{\pi NN}}{m^2_\pi}\,
V_{\pi NN}^{s'_2s_2}(\np'_2,\nh_2) 
F_{\pi N\Delta}(k_2^2)
\bar{u}_{s^\prime_1}(\np^\prime_1)
 k^\alpha_{2}
G_{\alpha\beta}(h_1+Q)
\Gamma^{\beta\mu}(Q)
u_{s_1}(\nh_1)
+(1\leftrightarrow2)
\label{deltaF}.
\\
j^\mu_{\Delta B}
&=&
\left[U_{\rm B}^{\pm}\right]_{1'2',12}\; 
\frac{f^* f_{\pi NN}}{m^2_\pi}\,
V_{\pi NN}^{s'_2s_2}(\np'_2,\nh_2) 
F_{\pi N\Delta}(k_2^2)
\bar{u}_{s^\prime_1}(\np^\prime_1)
k^\beta_{2}
\hat{\Gamma}^{\mu\alpha}(Q)
G_{\alpha\beta}(p^\prime_1-Q)
u_{s_1}(\nh_1)
+(1\leftrightarrow2)
\label{deltaB}.
\end{eqnarray}
where $k_i^\mu= (p'_i-h_i)^\mu$ is the four momentum transferred to
the $i$-th nucleon. Note that the energy of each nucleon includes the
vector energy, Eq. (\ref{energyrmf}), but the vector energies of the
initial and final nucleons cancel out when computing the components
$k_i^0=E'_i-E_i$.

In these equations we have defined the following function:
\begin{equation}
V_{\pi NN}^{s'_1s_1}(\np'_1,\nh_1) \equiv 
F_{\pi NN}(k_1^2)
\frac{\bar{u}_{s^\prime_1}(\np^\prime_1)\,\gamma_5
 \kbar_{1} \, u_{s_1}(\nh_1)}{k^2_{1}-m^2_\pi}.
\end{equation}
This function appears in all the currents, 
describing the propagation and emission (or absorption) of the exchanged pion,
having a strong form factor $F_{\pi NN}$ given by
\cite{Som78,Alb84}
\begin{equation}
F_{\pi NN}(k_1^2)= \frac{\Lambda^2-m_\pi^2}{\Lambda^2-k_1^2}
\end{equation}
and $\Lambda=1300$ MeV.

The charge dependence of the currents in the different p-n channels
 is determined by the matrix elements of the
  isospin operators $I_V^{\pm}$, $U_{\rm F}^{\pm}$, and   
$U_{\rm B}^{\pm}$, where the $+ (-)$ sign refers to 
 neutrino (antineutrino) scattering. They are defined by  
\begin{eqnarray}
I_V^{\pm}  &=& (I_V)_x\pm i (I_V)_y \\
U_{F}^{\pm} &=& (U_F)_x\pm i (U_F)_y\\
U_{B}^{\pm} &=& (U_B)_x\pm i (U_B)_y
\end{eqnarray}
Where
\begin{eqnarray}
\Ivec_V  & =&  i \left[\tauvec(1) \times\tauvec(2)\right] \\
(U_F)_j&=&\sqrt{\frac32}
\sum_{i=1}^3\left(T_i T_j^\dagger\right)\otimes
\tau_i
\label{forward}
\\
(U_B)_j &=&
\sqrt{\frac32}
\sum_{i=1}^3\left(T_{j}\, T^\dagger_i\right)\otimes
\tau_i,
\label{backward}
\end{eqnarray}
where $\vec{T}$ is an isovector transition
operator from isospin $\frac32$ to $\frac12$.

Additionally, each current contains coupling constants and electroweak
form factors. The coupling constants are: $f_{\pi NN}=1$, $g_A=1.26$
and  $f^*=2.13$. 

The electroweak form factors are $F_1^V=F_1^p-F_1^n$ in the seagull
vector and pion-in flight currents, for which we use Galster's
parametrization \cite{Gal71}, and $F_{\rho}$ in the axial seagull and
pion-pole currents, taken from \cite{Hernandez:2007qq}.  
In the case of the $\Delta$ current we use the
form factors $C_3^V$ and $C_5^A$, respectively, for the vector and
axial parts of the current \cite{Hernandez:2007qq}.  These appear in
the $N\rightarrow\Delta$ transition vertex in the forward current
\begin{equation} \label{gamma-delta}
\Gamma^{\beta\mu}(Q)=
\frac{C^V_3}{m_N}
\left(g^{\beta\mu}\Qbar-Q^\beta\gamma^\mu\right)\gamma_5
+ C^A_5 g^{\beta\mu},
\end{equation}
and for the backward current
\begin{equation}
\hat{\Gamma}^{\mu\alpha}(Q)=\gamma^0
\left[\Gamma^{\alpha\mu}(-Q)\right]^{\dagger}
\gamma^0 \, .
\end{equation}
The vector and axial form factors in the $\Delta$ current are from ref.
\cite{Hernandez:2007qq}
\begin{equation}
C_3^V(Q^2)= \frac{2.13}{(1-Q^2/M_V^2)^2}\frac{1}{1-\frac{Q^2}{4 M_V^2}},
\kern 1cm
C_5^A(Q^2)= \frac{1.2}{(1-Q^2/M_{A\Delta}^2)^2}\frac{1}{1-\frac{Q^2}{4 M_{A\Delta}^2}},
\end{equation}
with $M_V=0.84$ GeV, and $M_{A\Delta}=1.05$ GeV.

We use the $\pi N \Delta$ strong form factor of Ref. \cite{Dek94}, given by
\begin{equation}
F_{\pi N\Delta}(k_2^2)=
 \frac{\Lambda^2_\Delta}{\Lambda_\Delta^2-k_2^2}
\end{equation}
were $\Lambda_\Delta=1150$ MeV. 

Finally
the $\Delta$-propagator,
including the decay width of the $\Delta\,(1232)$,
is given by
\begin{equation}\label{delta_prop}
 G_{\alpha\beta}(P)= \frac{{\cal P}_{\alpha\beta}(P)}{P^2-
 M^2_\Delta+i M_\Delta \Gamma_\Delta(P^2)+
 \frac{\Gamma_{\Delta}(P^2)^2}{4}} \, .
\end{equation}
The projector  ${\cal P}_{\alpha\beta}(P)$ over 
spin-$\frac32$ on-shell particles is given by
\begin{eqnarray}
{\cal P}_{\alpha\beta}(P)&=&-(\Pbar+M_\Delta)
\left[g_{\alpha\beta}-\frac13\gamma_\alpha\gamma_\beta-
\frac23\frac{P_\alpha P_\beta}{M^2_\Delta}\right.
+\left.
\frac13\frac{P_\alpha\gamma_\beta-
P_\beta\gamma_\alpha}{M_\Delta}\right].
\end{eqnarray}

This concludes the description of the electroweak current.  The
spinors of particles and holes are calculated in the RMF model with
the relativistic effective mass $m_N^*$ instead of $m_N$, and with
on-shell energy, Eq. (\ref{onshell}). The total nucleon energy in the
RMF (\ref{energyrmf}) includes the vector energy, $E_v$. As can be
seen, this vector energy cancels out in the terms of the currents that
depend on the vectors $k_i^\mu$. But it is not canceled in the delta
propagator, which is the only place where $E_v$ will appear
explicitly.

Note that when considering the free $\Delta$ without self-nergy we are
neglecting the interaction of the $\Delta$ in the medium. 
The case of the interacting Delta is summarized in appendix \ref{apendice0}.
and is briefly discussed at the end of the results section.

The seagull current has a vector and axial part. The pionic current is
pure vector and the pion-pole current is pure axial. However, the
pion-pole could be considered as the axial part of the pion in flight
in a certain sense. The $\Delta$ forward and backward currents have a
vector and an axial part also.

In the case of electron scattering the above formulas are valid but
only the vector part of the current appears. Thus the pion-pole
diagrams (d,e) are not contributing. The isospin operators are modified
to the $z$ component instead of the $\pm$ component \cite{Mar21}.
Only the $L$ and $T$ responses are present in $(e,e')$

The exact calculation of the response functions $R_K(q,\omega)$ with
this model of currents is carried out by numerical integration in
seven dimensions of the Eq. (\ref{responses}). The spin sums are
performed numerically.  The isospin channels in the case of neutrino
CC scattering are $pn$ and $pp$ in the final states. These two
channels are computed separately and then added to obtain the total
responses. More details on the calculation are given in \cite{Sim17}.

Note that the same 2p2h responses were calculated in a
previous work \cite{Meg15,Meg16} with a similar MEC model.  The
differences with our new model are in the description of the $\Delta$
propagator, which is included here in its entirety, while in
\cite{Meg15,Meg16} only the real part of the denominator of the
$\Delta$ propagator was included. The second difference is that in the
present model the nucleon wave functions are solutions of the Dirac
equation in the mean field, that is, Dirac plane waves with
relativistic effective mass and vector energy. On the contrary in
references \cite{Meg15,Meg16} a RFG without effective mass was used,
but including an energy shift of 40 MeV.
As seen in the reference \cite{Mar21}, the effect of the imaginary part of the
propagator is to increase the contribution of the $\Delta F$ diagram and the
position of the maximum. The effect of the effective mass and vector
energy is to reduce the height of the peak, which partially
compensates for the difference of the MEC with respect to the
calculation of refs. \cite{Meg15,Meg16}.

Finally, the present model is restricted to a neutrino energy range in
which excitations of higher resonances, beyond the Delta, are not
essential. This is good enough for experiments like T2K, but not for
DUNE.

\subsection{Phase space integral}

Part of the dependence of the 2p2h response as a function of $q$ and
$\omega$ will be associated with the number of 2p2h states that can be
excited while conserving energy and momentum. That number is
proportional to the phase space function. In the RMF of nuclear
matter, it is given by
 \begin{eqnarray}
F(q,\omega)
&=&
\int
d^3p'_1
d^3h_1
d^3h_2
\frac{(m^*_N)^4}{E_1E_2E'_1E'_2}
\Theta(p'_1,h_1)\Theta(p'_2,h_2)
\delta(E'_1+E'_2-E_1-E_2-\omega) .
\label{espaciofasico}
\end{eqnarray}
where $\bf p'_2= h_1+h_2+q-p'_1$. 
This function was fully studied in
 \cite{Sim14,Sim14b}. 
The phase space determines
the global behavior of the 2p2h responses, on top of the additional
modifications introduced by the particular model of two-body current
operator. The main modification of the phase space behavior
in the responses is produced by the $Q^2$ dependence of the
electroweak form factors. The rest of the dependency on $(q,\omega)$
comes from the structure of the different
Feynman diagrams, which are dominated by the forward $\Delta$ excitations. 

In this work we approximate the phase space using the frozen nucleon approximation \cite{Sim14,Sim14b}, consisting in neglecting the momenta of the holes compared to the momentum transfer, which usually is larger than $k_F$. 
Setting $\nh_1=\nh_2=0$, and $E_1=E_2=m_N^*$ inside the integral 
(\ref{espaciofasico}) one can
perform the integral
over $\nh_1,\nh_2$. 
\begin{eqnarray}
F(q,\omega)_{\rm frozen}
= 
\left(\frac43\pi k_F^3\right)^2
\int d^3p'_1\;
\frac{(m^*_N)^2}{E'_1E'_2}\;
\Theta(p'_1,0)\Theta(p'_2,0) 
\delta(E'_1+E'_2-2m^*_N-\omega)\;
\label{fasico_frozen}
\end{eqnarray}
This integral can be reduced to one dimension, integrating over the
momentum $p'_1$ and the azimuthal angle $\phi'_1$ of the first ejected
nucleon. In ref. \cite{Sim14} the phase space function (\ref{espaciofasico}) 
was studied by
computing the 7-D integral numerically and compared to the frozen
approximation (\ref{fasico_frozen}). This approximation is very good even for low momenta,
except in the very low energy threshold zone, where the 2p2h cross
section is very small.

If we neglect Pauli Blocking, the integral in Eq.
 (\ref{fasico_frozen}) can be made
analytically by going to the center of mass frame of the two final
nucleons \cite{Sim14b}. The following approximate formula is obtained:
\begin{equation}
F(q,\omega)\simeq F(q,\omega)_{\rm frozen} \simeq
4\pi
\left( \frac{4}{3}\pi k_F^3 \right)^2
\frac{m_N^{* 2}}{2} \sqrt{1-\frac{4m_N^{*2}}{(2m_N^*+\omega )^2-q^2}}\,,
\label{analytical}
\end{equation}
From this equation
we obtain the minimum $\omega$  to excite a 2p2h state
 for fixed $q$, in this approximation. 
\begin{equation}
 \omega_{\rm min} =  \sqrt{4 m_N^{*2} + q^2} - 2 m^*_N
    \label{correctionf}
\end{equation}
The phase space integral is zero below this value.  Note that this is
the kinetic energy of a particle with mass $2m_N^*$ and momentum $q$.

\subsection{The averaged $\Delta$ propagator}

When a $\Delta$ is excited in the $\Delta$ forward diagrams (f) and (g) of Fig
\ref{fig_feynman} a broad resonant peak is produced. This produces the
dominant contribution to the MEC responses in the region around the
$\Delta$ peak.  To get this resonant peak in the semi-empirical formula
we will approximate it by an average of the $\Delta$ propagator over the Fermi
gas.

The peak of the 2p2h responses is due to the 
denominator in the the $\Delta$ propagator for the forward diagrams
\begin{equation}
G_{\Delta}(H+Q)
\equiv
\frac{1}{(H+Q)^2-
 M^2_\Delta+i M_\Delta \Gamma_\Delta+
 \frac{\Gamma^2_{\Delta}}{4}} \, ,
\label{denominator}
\end{equation}
where $H^\mu=(E_{RMF},\nh)$ is the four-momentum of the hole. 
The square of the momentum in the denominator can be written
\begin{equation}
(H+Q)^2  
= (E_{RMF}+\omega)^2-(\nh+\nq)^2 
\simeq  (m^*_N+E_v+\omega)^2-2\nh\cdot\nq-q^2 
\end{equation}
where the energy of the hole has been approximated by $E_{RMF}=m^*_N+E_v$ in
the RMF model, by neglecting the kinetic energy of the first particle
and the momentum $h^2$ has also been neglected compared to the
momentum transfer $q^2$. In this non relativistic limit for the initial
nucleon, an average of the propagator (\ref{denominator}) can be
computed analytically
\begin{eqnarray}
G_{\rm av}(Q)=G_{\rm av}(q,\omega)
&=&
\frac{1}{\frac43 \pi k^3_F}
\int 
 \frac{d^3h \;\theta(k_F-\left|\nh\right|)}{a-2\,\nh\cdot\nq+ib} \, ,
\\
&=& 
\frac{1}{\frac43 \pi k^3_F} 
\frac{\pi}{q}
\left\{ \frac{\left(a+ib\right)k_F}{2q}
+ \frac{4q^2k^2_F - (a+ib)^2}{8q^2}
\ln\left[\frac{a+2k_Fq+ib}{a-2k_Fq+ib}\right]
\right\} \, ,
\label{averaged_denom}
\end{eqnarray}
where the functions $a,b$ are defined by  
\begin{eqnarray}
\kern -0.8cm 
a &\equiv& m^{*2}_N+(\omega+E_v)^2-q^2+2m^*_N(\omega+E_v+\Sigma)-M^2_\Delta+
\frac{\Gamma^2}{4}
\label{afrozen}\\
\kern -0.8cm 
b &\equiv& M_\Delta \Gamma
\label{bfrozen} \,.
\end{eqnarray}
Note that we have included a shift 
parameter $\Sigma(q)$, 
to obtain the correct position of the smeared $\Delta$ peak.
The $\Delta$ width  $\Gamma_\Delta$ will be  replaced by  an effective width
on the average $\Gamma(q)$.  

The effective width and shift are used only in the averaged
propagator.  In the exact calculation of the MEC there is no shift and
the well-known value of the width of the Delta 
$ \Gamma_\Delta (Q^2) $ 
is used.  In reference \cite{Sim17b} it was shown that the
averaged propagator (in \cite{Sim17b} it was called ``frozen'' propagator) 
describes the $\omega$-dependence of the 2p2h
responses only if effective values of $\Gamma(q)$ and $\Sigma(q)$ are
used. The effective values are taken as parameters in the
semi-empirical formula and they are fitted to the exact responses.  As
explained in detail in \cite{Sim17b}, in the exact responses the
$\Delta$ propagator inside the 7D-integral is being multiplied by a
q-dependent weight determined by the matrix elements of the MEC. This
changes the position and width of the MEC peak with respect to the
simple average of the denominator.

The effective shift and width parameters, $\Sigma(q)$, and
$\Gamma(q)$, will be adjusted with the semi-empirical formula of next
section, for each value of the momentum transfer $q$.

\section{The semi-empirical formulas of MEC responses}

In this section we propose 
the semi-empirical formulas for the nuclear responses,
$R_K(q,\omega)$, by separating the contributions of the different
Feynman diagrams of MEC and the interferences between them, and
extracting the contribution from the phase space, the electroweak form
factors and coupling constants, corresponding to each term of the
current. In the case of the $\Delta$ forward current we also extract
an average value of the $\Delta$ propagator. Much of the dependence on
$q$ and $\omega$ is coming from these factors. The remaining
dependence is coded into coefficients $\tilde{C}_i(q)$ that are
assumed to only depend on $q$.

First we write the MEC as
\begin{equation}
j^\mu = j_{SP}^\mu +j_{\Delta F}^\mu +j_{\Delta B}^\mu
\end{equation}
where the seagull-pionic (SP) contribution is the sum of diagrams (a--e) of Fig.
\ref{fig_feynman}
\begin{equation}
j_{SP}^\mu = j_{\rm sea}^\mu +j_{\pi}^\mu +j_{\rm pole}^\mu
\end{equation}
Now to compute the hadronic tensor $w^{\mu\nu}$,
Eq. (\ref{elementary}), we deal with products of the kind
\begin{equation}
j^{\mu *}j^\nu  = j^{\mu *}_{SP}j^\nu_{SP}  +
                j^{\mu *}_{\Delta F}j^\nu_{\Delta F} +  
                j^{\mu *}_{\Delta B}j^\nu_{\Delta B}  + 
                j^{\mu *}_{SP}j^\nu_{\Delta F}+   
                j^{\mu *}_{\Delta F}j^\nu_{SP}   +
                j^{\mu *}_{SP}j^\nu_{\Delta B}+   
                j^{\mu *}_{\Delta B}j^\nu_{SP}+   
                j^{\mu *}_{\Delta F}j^\nu_{\Delta B}+   
                j^{\mu *}_{\Delta B}j^\nu_{\Delta F}   
\end{equation}
Using this expansion the response functions (\ref{rcc}--\ref{rtprima}) can be written as the sum of six 
sub-responses corresponding to SP, $\Delta F$, $\Delta B$, 
plus the interferences $\Delta F-SP$, $\Delta B-SP$, and $\Delta F-\Delta B$
\begin{equation}
R^{K}(q,\omega) = R^K_{SP} + R^K_{\Delta F}+ R^K_{\Delta B} +
R^K_{\Delta F-SP} +R^K_{\Delta B-SP} +R^K_{\Delta F-\Delta B}
\end{equation}
this is a general expansion for all responses.  In the case of the
$CC, CL, LL$ and $T$ responses we can also separate the contribution
of the vector and axial part of the current, 
\begin{equation}
R^K(q,\omega) =  R^{K,VV} + R^{K,AA},    \kern 1cm K=CC, CL, LL, T 
\end{equation}
and their expansion is
\begin{eqnarray}
R^{K,VV}(q,\omega) &=& R^{K,VV}_{SP} + R^{K,VV}_{\Delta F}+ R^{K,VV}_{\Delta B} +
R^{K,VV}_{\Delta F-SP} +R^{K,VV}_{\Delta B-SP} +R^{K,VV}_{\Delta F-\Delta B}
\\
R^{K,AA}(q,\omega) &=& R^{K,AA}_{SP} + R^{K,AA}_{\Delta F}+ R^{K,AA}_{\Delta B} +
R^{K,AA}_{\Delta F-SP} +R^{K,AA}_{\Delta B-SP} +R^{K,AA}_{\Delta F-\Delta B}
\end{eqnarray}
In the case of the $T'$ response, 
only the vector-axial product contributes, 
\begin{equation}
R^{T'}(q,\omega) =
R^{T',VA}(q,\omega) = R^{T',VA}_{SP} + R^{T',VA}_{\Delta F}+ R^{T',VA}_{\Delta B} +
R^{T',VA}_{\Delta F-SP} +R^{T',VA}_{\Delta B-SP} +R^{T',VA}_{\Delta F-\Delta B}
\end{equation}
We have written the 2p2h response functions as sums of the
sub-responses vector-vector,  $R^{K,VV}_{I,J}$, axial-axial, $R^{K,AA}_{I,J}$, and
vector-axial, $R^{T',VA}_{I,J}$, with $ I,J= SP, \Delta F, \Delta B$. From each one
of these sub-responses we factorize the electroweak form factors, the 
coupling constants, an average delta propagator $G_{\rm av}(q,\omega)$
for each $\Delta F$ current, and, finally,
 the phase-space  $\frac{V}{(2\pi)^9} F(q,\omega)$.
Then for each sub-response we propose a semi-empirical formula. Schematically
the general structure will be 
\begin{equation}
R_i(q,\omega) = [\mbox{phase-space}] 
\times [\mbox{coupling constants}]  
\times [\mbox{form factors}]  
\times [\mbox{averaged $\Delta$ propagators}]  
\times \tilde{C}_i(q)
\end{equation}
where we assume that the adjustable coefficients $\tilde{C}_i(q)$ do
not depend on $\omega$, but only depend on $q$. 
This is the main hypothesis on which the parametrization is based,
that is, that most of the $\omega$-dependence comes from phase space,
form factors and the averaged $\Delta$ propagator. This is justified
a posteriori in the next section when we check the quality of the fit
by comparison with the exact results. These
coefficients will be fitted to the corresponding sub-responses in an
exact calculation.  The coefficients can be interpreted as nuclear
mean values of spin-isospin contributions of the Feynman diagrams for
each sub-response in a 2p2h excitation.

Below we write down the explicit formula for the 54 sub-responses, taking
into account that some of them may need two coefficients or some
additional correction, which will be discussed in the next section.

\subsection{Response $R_T^{VV}$}

\begin{equation}
R_{\Delta F}^{T,VV}
=
\frac{V}{(2\pi)^9}
F(q,\omega)
\left( \frac{f^*f_{\pi NN}}{m_{\pi}^2}\right)^2
\left( \frac{C^V_3}{m_N} \right)^2
\left[ \tilde{C}_{1,V1}(Re(G^V_{\rm av}))^2  +  \tilde{C}_{1,V2}  (Im(G^V_{\rm av}))^2 \right]
 (m_N^4)
\label{rtforv}
\end{equation}

\begin{equation}
R_{SP}^{T,VV}
=
\frac{V}{(2\pi)^9}
F(q,\omega)
\left( \frac{f_{\pi NN}^2}{m_{\pi}^2}\right)^2
\left( F^V_1 \right)^2
(\tilde{C}_{2,V} \cdot m_N^{-2}) 
\left [1-\frac{\omega-0.7q}{m_N} \right]^2
\label{rtspv}
\end{equation}
\begin{equation}
R_{\Delta B}^{T,VV}
=
\frac{V}{(2\pi)^9}
F(q,\omega)
\left( \frac{f^*f_{\pi NN}}{m_{\pi}^2}\right)^2
\left( \frac{C^V_3}{m_N} \right)^2
(\tilde{C}_{3,V})
\label{rtbackv}
\end{equation}
\begin{equation}
R_{\Delta F - SP}^{T,VV}
=
\frac{V}{(2\pi)^9}
F(q,\omega)
\left( \frac{f^*f_{\pi NN}^3}{m_{\pi}^4}\right)
\left( \frac{C^V_3}{m_N} \right)
\left( F^V_1 \right)
\left[ \tilde{C}_{4,V1}(Re(G^V_{\rm av}))  +  \tilde{C}_{4,V2}  (Im(G^V_{\rm av})) \right]
(m_N)
\label{rtforspv}
\end{equation}
\begin{equation}
R_{\Delta F - \Delta B}^{T,VV}
=
\frac{V}{(2\pi)^9}
F(q,\omega)
\left( \frac{f^*f_{\pi NN}}{m_{\pi}^2}\right)^2
\left( \frac{C^V_3}{m_N} \right)^2
\left[ \tilde{C}_{5,V1}(Re(G^V_{\rm av}))  +  \tilde{C}_{5,V2} (Im(G^V_{\rm av})) \right]
(m_N^2)
\label{rtforbackv}
\end{equation}
\begin{equation}
R_{\Delta B - SP}^{T,VV}
=
\frac{V}{(2\pi)^9}
F(q,\omega)
\left( \frac{f^*f_{\pi NN}^3}{m_{\pi}^4}\right)
\left( \frac{C^V_3}{m_N} \right)
\left( F^V_1 \right)
(\tilde{C}_{6,V} \cdot m_N ^{-1})
\label{rtbackspv}
\end{equation}

\subsection{Response $R_T^{AA}$}

\begin{equation}
R_{\Delta F}^{T,AA}
=
\frac{V}{(2\pi)^9}
F(q,\omega)
\left( \frac{f^*f_{\pi NN}}{m_{\pi}^2}\right)^2
\left( C^5_A \right)^2
\left[ \tilde{C}_{1,A1}(Re(G^A_{\rm av}))^2  +  \tilde{C}_{1,A2}  (Im(G^A_{\rm av}))^2 \right]
( m_N^2)
\label{rtfora}
\end{equation}
\begin{equation}
R_{SP}^{T,AA}
=
\frac{V}{(2\pi)^9}
F(q,\omega)
\left( \frac{f_{\pi NN}^2}{m_{\pi}^2}\right)^2
\left( \frac{1}{g_A} \right)^2
(\tilde{C}_{2,A} \cdot m_N^{-2})
\label{rtspa}
\end{equation}
\begin{equation}
R_{\Delta B}^{T,AA}
=
\frac{V}{(2\pi)^9}
F(q,\omega)
\left( \frac{f^*f_{\pi NN}}{m_{\pi}^2}\right)^2
\left( C^A_5 \right)^2
(\tilde{C}_{3,A} \cdot m_N^{-2})
\label{rtbacka}
\end{equation}
\begin{equation}
R_{\Delta F - SP}^{T,AA}
=
\frac{V}{(2\pi)^9}
F(q,\omega)
\left( \frac{f^*f_{\pi NN}^3}{m_{\pi}^4}\right)
\left( C^A_5  \right)
\left( \frac{1}{g_A} \right)
\left[ \tilde{C}_{4,A1}(Re(G^A_{\rm av}))  +  \tilde{C}_{4,A2}  (Im(G^A_{\rm av})) \right]
\label{rtforspa}
\end{equation}
\begin{equation}
R_{\Delta F - \Delta B}^{T,AA}
=
\frac{V}{(2\pi)^9}
F(q,\omega)
\left( \frac{f^*f_{\pi NN}}{m_{\pi}^2}\right)^2
\left( C^A_5  \right)^2
\left[ \tilde{C}_{5,A1}(Re(G^A_{\rm av}))  +  \tilde{C}_{5,A2}  (Im(G^A_{\rm av})) \right]
\label{rtforbacka}
\end{equation}

\begin{equation}
R_{\Delta B - SP}^{T,AA}
=
\frac{V}{(2\pi)^9}
F(q,\omega)
\left( \frac{f^*f_{\pi NN}^3}{m_{\pi}^4}\right)
\left( C^A_5  \right)
\left( \frac{1}{g_A} \right)
(\tilde{C}_{6,A} \cdot m_N ^{-2})
\label{rtbackspa}
\end{equation}

\subsection{Response $R_{T'}^{VA}$}

\begin{equation}
R_{\Delta F}^{T',VA}
=
\frac{V}{(2\pi)^9}
F(q,\omega)
\left( \frac{f^*f_{\pi NN}}{m_{\pi}^2}\right)^2
\left( \frac{C^V_3}{m_N} \right)
\left( C^A_5  \right)
\left[ \tilde{C}_{1,VA1}(Re(G^{VA}_{\rm av}))^2  +  \tilde{C}_{1,V
A2}  (Im(G^{VA}_{\rm av}))^2 \right]
 (m_N^3)
\label{rtpfor}
\end{equation}
\begin{equation}
R_{SP}^{T',VA}
=
\frac{V}{(2\pi)^9}
F(q,\omega)
\left( \frac{f_{\pi NN}^2}{m_{\pi}^2}\right)^2
\left( F^V_1 \right)
\left( \frac{1}{g_A} \right)
(\tilde{C}_{2,VA} \cdot m_N^{-2})\left [1-\frac{\omega-0.7q}{m_N} \right]^2
\label{rtpsp}
\end{equation}
\begin{equation}
R_{\Delta B}^{T',VA}
=
\frac{V}{(2\pi)^9}
F(q,\omega)
\left( \frac{f^*f_{\pi NN}}{m_{\pi}^2}\right)^2
\left( \frac{C^V_3}{m_N} \right)
\left( C^A_5  \right)
(\tilde{C}_{3,VA} \cdot m_N ^{-1})
\label{rtpback}
\end{equation}
\begin{equation}
R_{\Delta F - SP}^{T',VA}
=
\frac{V}{(2\pi)^9}
F(q,\omega)
\left( \frac{f^*f_{\pi NN}^3}{m_{\pi}^4}\right)
\left| G^{VA}_{\rm av} \right|
\left [
\left( \frac{C^V_3}{m_N} \right)
\left( \frac{1}{g_A} \right)
(\tilde{C}_{4,VA1} \cdot m_N)
+
\left( C^A_5  \right)
\left( F^V_1 \right)
(\tilde{C}_{4,VA2})
\right ]
\label{rtpforsp}
\end{equation}
\begin{equation}
R_{\Delta F - \Delta B}^{T',VA}
=
\frac{V}{(2\pi)^9}
F(q,\omega)
\left( \frac{f^*f_{\pi NN}}{m_{\pi}^2}\right)^2
\left( \frac{C^V_3}{m_N} \right)
\left( C^A_5  \right)
\left[ \tilde{C}_{5,VA1}(Re(G^{VA}_{\rm av}))  +  \tilde{C}_{5,VA2} (Im(G^{VA}_{\rm av})) \right]
2 m_N
\label{rtpforback}
\end{equation}

\begin{equation}
R_{\Delta B - SP}^{T',VA}
=
\frac{V}{(2\pi)^9}
F(q,\omega)
\left( \frac{f^*f_{\pi NN}^3}{m_{\pi}^4}\right)
\left [
\left( \frac{C^V_3}{m_N} \right)
\left( \frac{1}{g_A} \right)
(\tilde{C}_{6,VA} \cdot m_N^{-1})
+
\left( C^A_5  \right)
\left( F^V_1 \right)
(\tilde{C}_{6,AV} \cdot m_N^{-2})
\right ]
\label{rtpbacksp}
\end{equation}

\subsection{Response $R_{CC}^{VV}$}

\begin{equation}
R_{\Delta F}^{CC,VV}
=
\frac{V}{(2\pi)^9}
F(q,\omega)
\left( \frac{f^*f_{\pi NN}}{m_{\pi}^2}\right)^2
\left( \frac{C^V_3}{m_N} \right)^2
\left[ \tilde{C}_{1,V1}(Re(G^V_{\rm av}))^2  +  \tilde{C}_{1,V2}  (Im(G^V_{\rm av}))^2 \right]
 (m_N^4)
\label{rccforv}
\end{equation}

\begin{equation}
R_{SP}^{CC,VV}
=
\frac{V}{(2\pi)^9}
F(q,\omega)
\left( \frac{f_{\pi NN}^2}{m_{\pi}^2}\right)^2
\left( F^V_1 \right)^2
(\tilde{C}_{2,V} \cdot m_N^{-2}) 
\label{rccspv}
\end{equation}

\begin{equation}
R_{\Delta B}^{CC,VV}
=
\frac{V}{(2\pi)^9}
F(q,\omega)
\left( \frac{f^*f_{\pi NN}}{m_{\pi}^2}\right)^2
\left( \frac{C^V_3}{m_N} \right)^2
(\tilde{C}_{3,V})
\label{rccbackv}
\end{equation}

\begin{equation}
R_{\Delta F - SP}^{CC,VV}
=
\frac{V}{(2\pi)^9}
F(q,\omega)
\left( \frac{f^*f_{\pi NN}^3}{m_{\pi}^4}\right)
\left( \frac{C^V_3}{m_N} \right)
\left( F^V_1 \right)
\left[ \tilde{C}_{4,V1}(Re(G^V_{\rm av}))  +  \tilde{C}_{4,V2}  (Im(G^V_{\rm av})) \right]
(m_N)
\label{rccforspv}
\end{equation}

\begin{equation}
R_{\Delta F - \Delta B}^{CC,VV}
=
\frac{V}{(2\pi)^9}
F(q,\omega)
\left( \frac{f^*f_{\pi NN}}{m_{\pi}^2}\right)^2
\left( \frac{C^V_3}{m_N} \right)^2
\left[ \tilde{C}_{5,V1}(Re(G^V_{\rm av}))  +  \tilde{C}_{5,V2} (Im(G^V_{\rm av})) \right]
(m_N^2)
\label{rccforbackv}
\end{equation}

\begin{equation}
R_{\Delta B - SP}^{CC,VV}
=
\frac{V}{(2\pi)^9}
F(q,\omega)
\left( \frac{f^*f_{\pi NN}^3}{m_{\pi}^4}\right)
\left( \frac{C^V_3}{m_N} \right)
\left( F^V_1 \right)
(\tilde{C}_{6,V} \cdot m_N ^{-1})
\label{rccbackspv}
\end{equation}

\subsection{Response $R_{CC}^{AA}$}
\begin{equation}
R_{\Delta F}^{CC,AA}
=
\frac{V}{(2\pi)^9}
F(q,\omega)
\left( \frac{f^*f_{\pi NN}}{m_{\pi}^2}\right)^2
\left( C^5_A \right)^2
\left[ \tilde{C}_{1,A1}(Re(G^A_{\rm av}))^2  +  \tilde{C}_{1,A2}  (Im(G^A_{\rm av}))^2 \right]
( m_N^2)
\label{rccfora}
\end{equation}
\begin{equation}
R_{SP}^{CC,AA}
=
\frac{V}{(2\pi)^9}
F(q,\omega)
\left( \frac{f_{\pi NN}^2}{m_{\pi}^2}\right)^2
\left( \frac{1}{g_A} \right)^2
\left [\tilde{C}_{2,A1} +  \tilde{C}_{2,A2} \left (\frac{\omega \cdot m_N}{Q^2-m_\pi^2} \right)^2
\right ]
( m_N^{-2})
\label{rccspa}
\end{equation}
\begin{equation}
R_{\Delta B}^{CC,AA}
=
\frac{V}{(2\pi)^9}
F(q,\omega)
\left( \frac{f^*f_{\pi NN}}{m_{\pi}^2}\right)^2
\left( C^A_5 \right)^2
(\tilde{C}_{3,A} \cdot m_N^{-2})
\label{rccbacka}
\end{equation}
\begin{equation}
R_{\Delta F - SP}^{CC,AA}
=
\frac{V}{(2\pi)^9}
F(q,\omega)
\left( \frac{f^*f_{\pi NN}^3}{m_{\pi}^4}\right)
\left( C^A_5  \right)
\left( \frac{1}{g_A} \right)
\left[ \tilde{C}_{4,A1}(Re(G^A_{\rm av}))  +  \tilde{C}_{4,A2}  (Im(G^A_{\rm av})) \right]
\label{rccforspa}
\end{equation}
\begin{equation}
R_{\Delta F - \Delta B}^{CC,AA}
=
\frac{V}{(2\pi)^9}
F(q,\omega)
\left( \frac{f^*f_{\pi NN}}{m_{\pi}^2}\right)^2
\left( C^A_5  \right)^2
\left[ \tilde{C}_{5,A1}(Re(G^A_{\rm av}))  +  \tilde{C}_{5,A2}  (Im(G^A_{\rm av})) \right]
\label{rccforbacka}
\end{equation}

\begin{equation}
R_{\Delta B - SP}^{CC,AA}
=
\frac{V}{(2\pi)^9}
F(q,\omega)
\left( \frac{f^*f_{\pi NN}^3}{m_{\pi}^4}\right)
\left( C^A_5  \right)
\left( \frac{1}{g_A} \right)
(\tilde{C}_{6,A} \cdot m_N ^{-2})
\label{rccbackspa}
\end{equation}

\subsection{Responses $R_{CL}^{VV}$ and $R_{LL}^{VV}$}

These responses are computed assuming conservation of the vector current

\begin{eqnarray}
R^{CL,VV} &=& - \frac{\omega}{q}  R^{CC,VV}\\
R^{LL,VV} &=& \frac{\omega^2}{q^2}  R^{CC,VV}
\end{eqnarray}

\subsection{Responses $R_{CL}^{AA}$ and $R_{LL}^{AA}$}

The semi-empirical formulas for $R_{CL}^{AA}$ and $R_{LL}^{AA}$ are
similar to the $R_{CC}^{AA}$. Only the numerical values of the
coefficients $\tilde{C}_i$ change.

\subsection{Electromagnetic responses $R_{em}^{L}$ and $R_{em}^{T}$}

It can be shown with the formalism of ref. \cite{Sim17}, that for
symmetric nuclear matter, the electromagnetic 2p2h responses are one
half of the $VV$ weak responses
\begin{eqnarray}
R_{em}^{L} &=& \frac{1}{2} R^{CC,VV}\\ 
R_{em}^{T} &=& \frac{1}{2} R^{T,VV}\\
\end{eqnarray}
Therefore the same semi-empirical formulas for the VV responses apply for the
electromagnetic responses with a factor 1/2.



\subsection{The properties of semi-empirical formulas}

Here we describe and clarify some particularities about the semi-empirical 
formulas
(\ref{rtforv}--\ref{rccbackspa})
for the sub-responses,
and their theoretical meaning.

\begin{itemize}
\item All the dependence on $\omega$ is analytical.  So the
  semi-empirical expansion assumes that the $\omega$ dependence comes
  mainly from the product of phase space $F(q,\omega)$, electroweak
  form factors and averaged $\Delta$ propagator. The only exceptions
  are $R^{T,VV}_{SP}$, and $R^{T',VA}_{SP}$ sub-responses, which
  include an $\omega$-dependent factor that is obtained empirically by comparing with the exact result.
 
\item The coefficients $\tilde{C}_i$ in all the formulas are
  dimensionless. That is why powers of nucleon masses have been
  introduced in the sub-responses.

\item The phase space $F(q,\omega)$ is common to all the formulas. In this work 
it is computed analytically using the approximation (\ref{analytical}).

\item All the responses are proportional to the volume
$V= (2\pi)^3 Z/(\frac8 3 \pi k_F^3)$ for symmetric nuclear matter $Z=N$.
For asymmetric matter, the
formulas should be modified using two different Fermi momenta
for protons and neutrons.

\item Each sub-response includes an specific product of form factors
  and coupling constants, except the  axial $SP$ subresponses, that do
  not allow to extract explicitly the form factor $F_\rho(k_i)$.

\item In most sub-responses the averaged $\Delta$ propagator appears
  separated in real and imaginary parts, each one multiplied by a
  parameter $\tilde{C}_i$.  The only exception is the $R^{T',VA}_{\Delta F-SP}$,
  that only include the modulus $|G_{\rm av}|$.  Note that in the
  formulas there are three versions of the averaged propagator:
  $G^V_{\rm av}$ for the $VV$ responses, $G^A_{\rm av}$ for the $AA$
  sub-responses and , $G^{VA}_{\rm av}$ for the $T'$, $VA$ responses.  They
  differ in the values of the effective width, $\Gamma$, and shift,
  $\Sigma$, of the $\Delta$ propagator. 
The corresponding six parameters are denoted by 
 $\Gamma_V$, $\Gamma_A$,
  $\Gamma_{VA}$,
 $\Sigma_V$,
  $\Sigma_A$, $\Sigma_{VA}$.

\end{itemize}

Finally, the semi-empirical formulas allow to calculate analytically
and directly all the 2p2h electro-weak responses.  For a fixed value
of $q$ the total 2p2h response depends on the sum of all the
sub-responses, with a total of 73 parameters. This number of
parameters may seem large. However, note that the five separate
responses are being described simultaneously, as well as their axial
and vector parts all together. This could be compared with the
parameterization of ref. \cite{Meg15,Meg16} which needs about 56
parameters to describe all the 2p-2h responses in another nuclear
model, based on the RFG and not RMF, and also using only the real part
of the $\Delta$ propagator. Here we use the full $\Delta$ propagator
\cite{Mar21}.  However the present parameterization is an advance
 since we obtain explicit dependence of
the responses on physical magnitudes ---form factors, coupling
constants, Fermi momentum, etc--- that can be modified a posteriori if
desired. Also as we show below many of these 54 sub-responses are very small
and could safely neglected,
leaving us with a smaller number of parameters.
However in this paper we have computed all the sub-responses.

\section{Results}

\begin{figure*}
\includegraphics[width=14cm, bb=31 273 534 780]{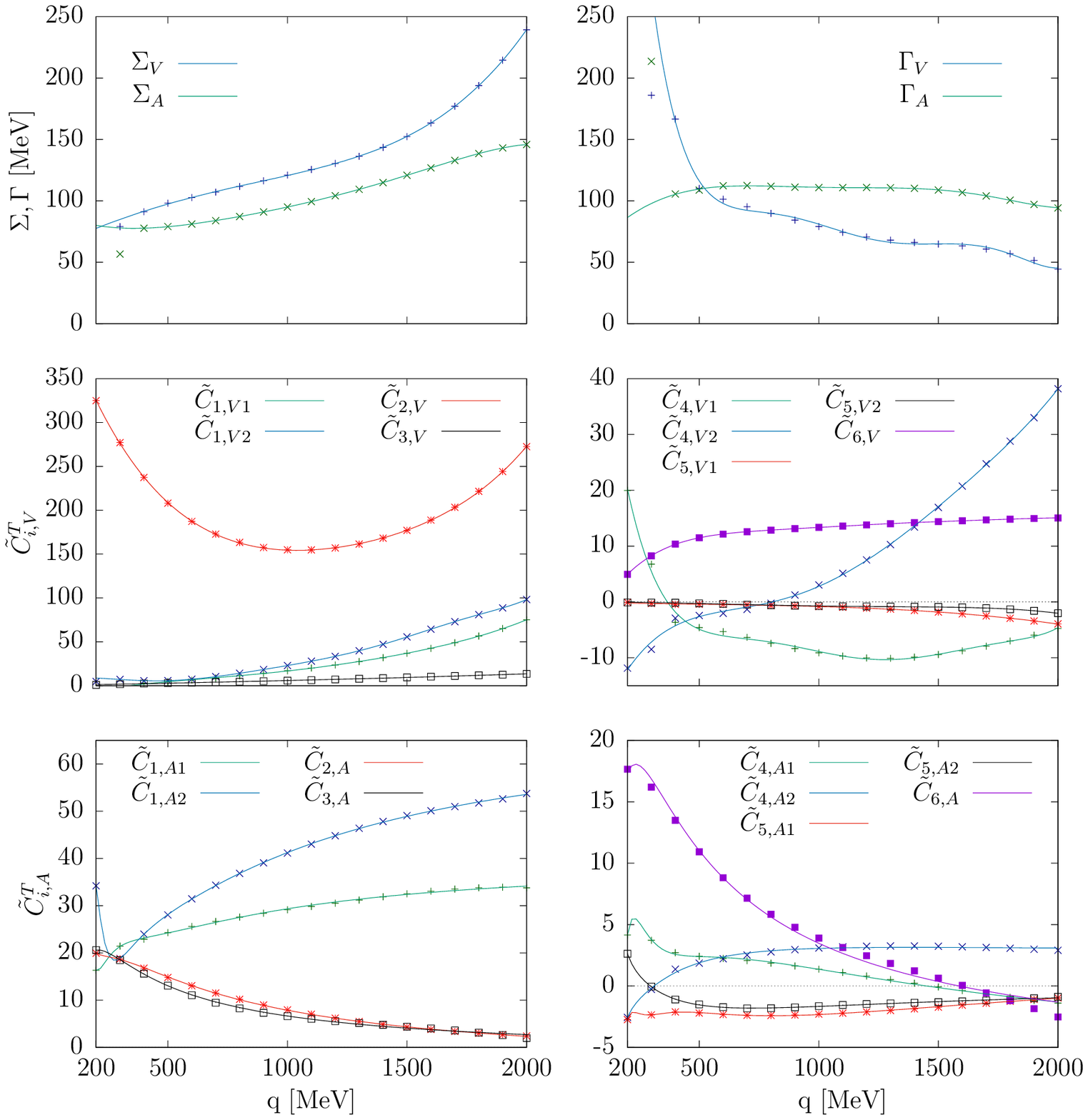}
\caption{ Coefficients of the semi-empirical formulas and parameters
  $\Gamma$, $\Sigma$ in the averaged $\Delta$ propagator, for the
  $R^T$ 2p2h response function plotted against the momentum transfer
  $q$. The dots are the fitted values and the curves are the
  parametrizations with polynomial functions. Tables of these
  coefficients and parametrizations are given in the Appendix.  }
\label{revalida1}
\end{figure*}

\begin{figure*}
\includegraphics[width=16cm, bb=31 173 534 680]{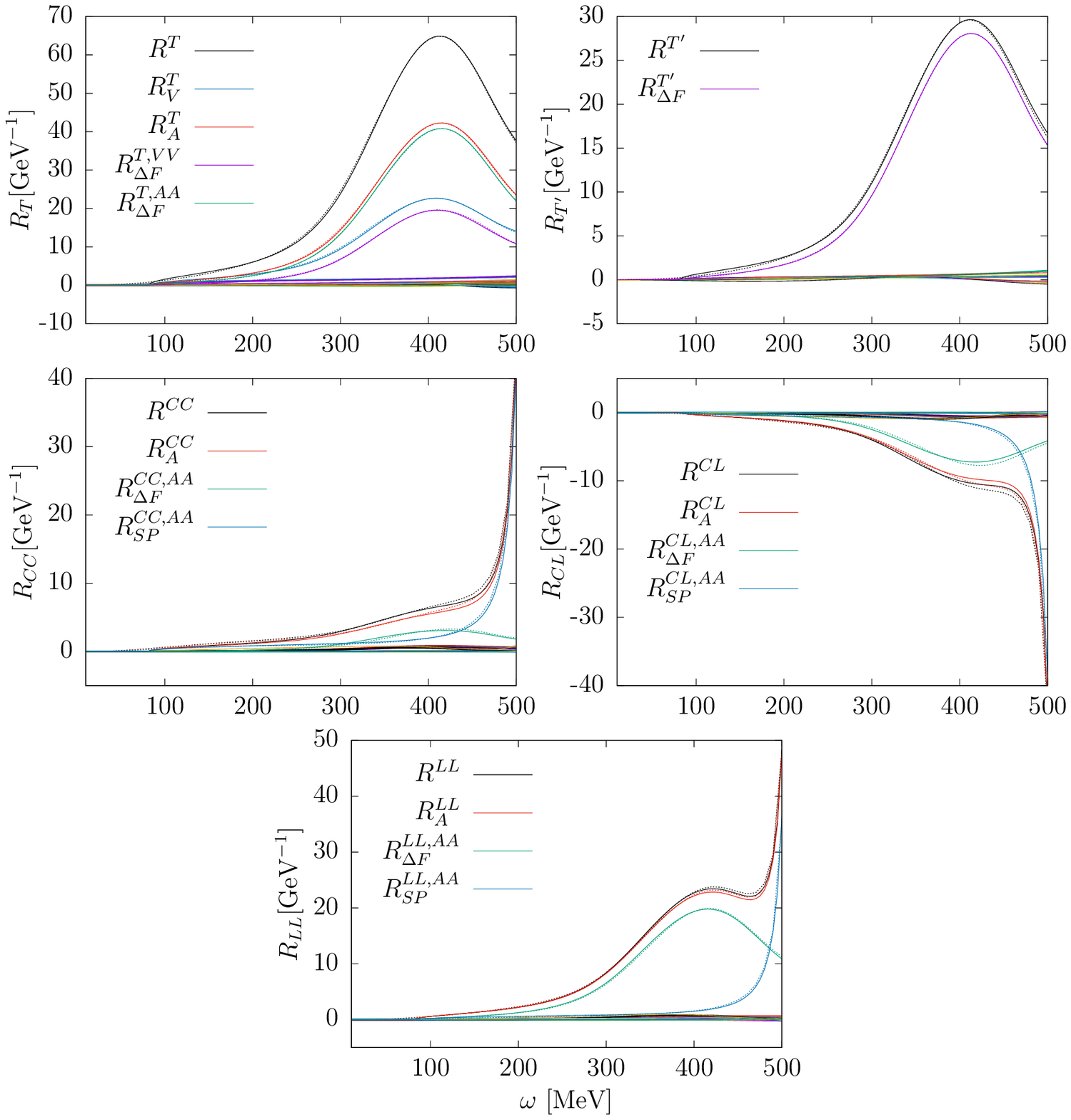}
\caption{
Comparison between all the sub-responses that contribute to each of the response functions for $q = 500$ MeV/c. Only the most important sub-responses appear in the legend. But they are all drawn. For each color the solid line is the parameterization and the dotted lines is the exact calculation.
 }
\label{revalida2a}
\end{figure*}

\begin{figure*}
\includegraphics[width=16cm, bb=31 173 534 680]{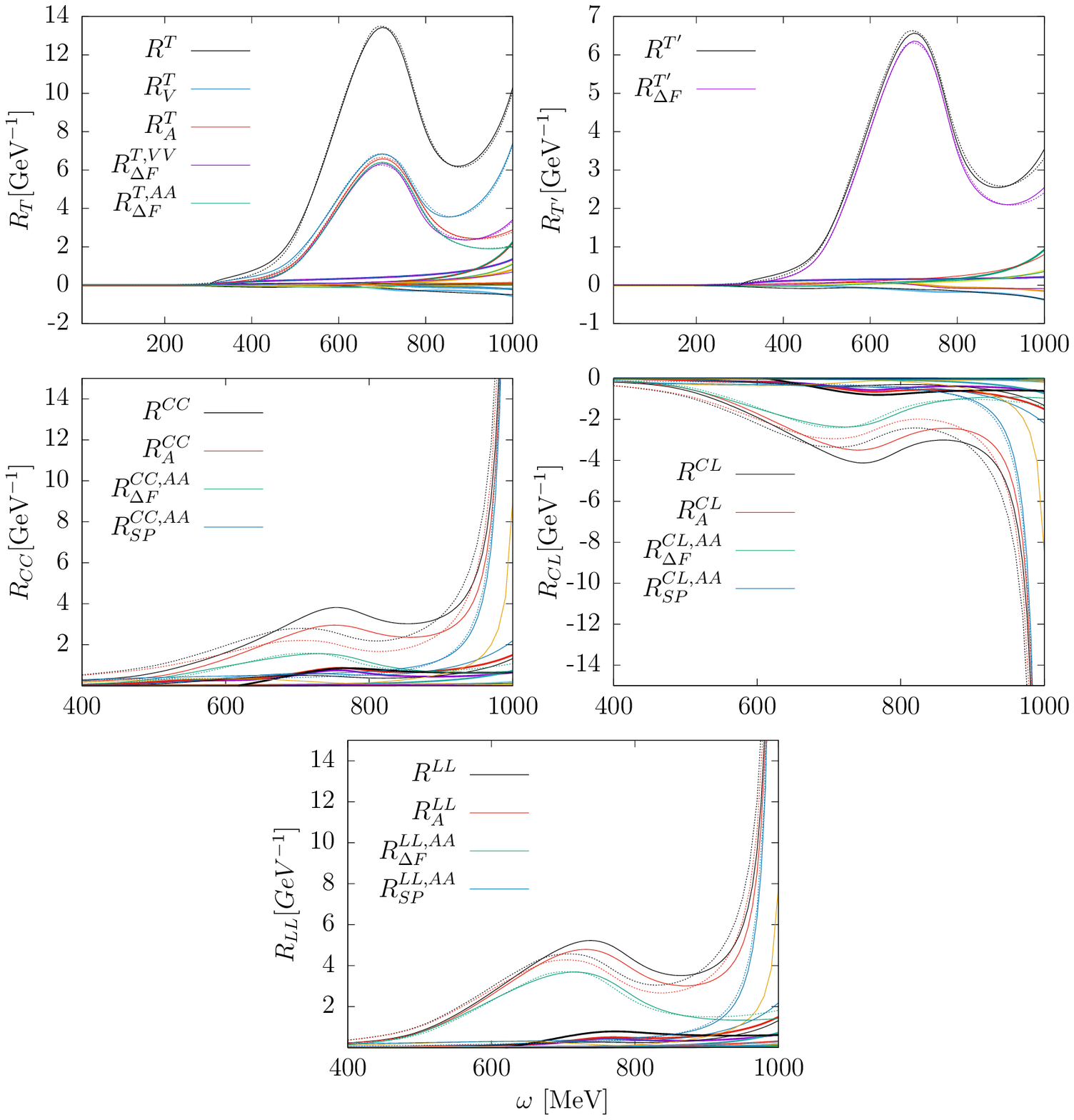}
\caption{
The same as Fig. \ref{revalida2a} for $q=1000$ MeV/c.
 }
\label{revalida2b}
\end{figure*}

\begin{figure*}
\includegraphics[width=16cm, bb=31 173 534 680]{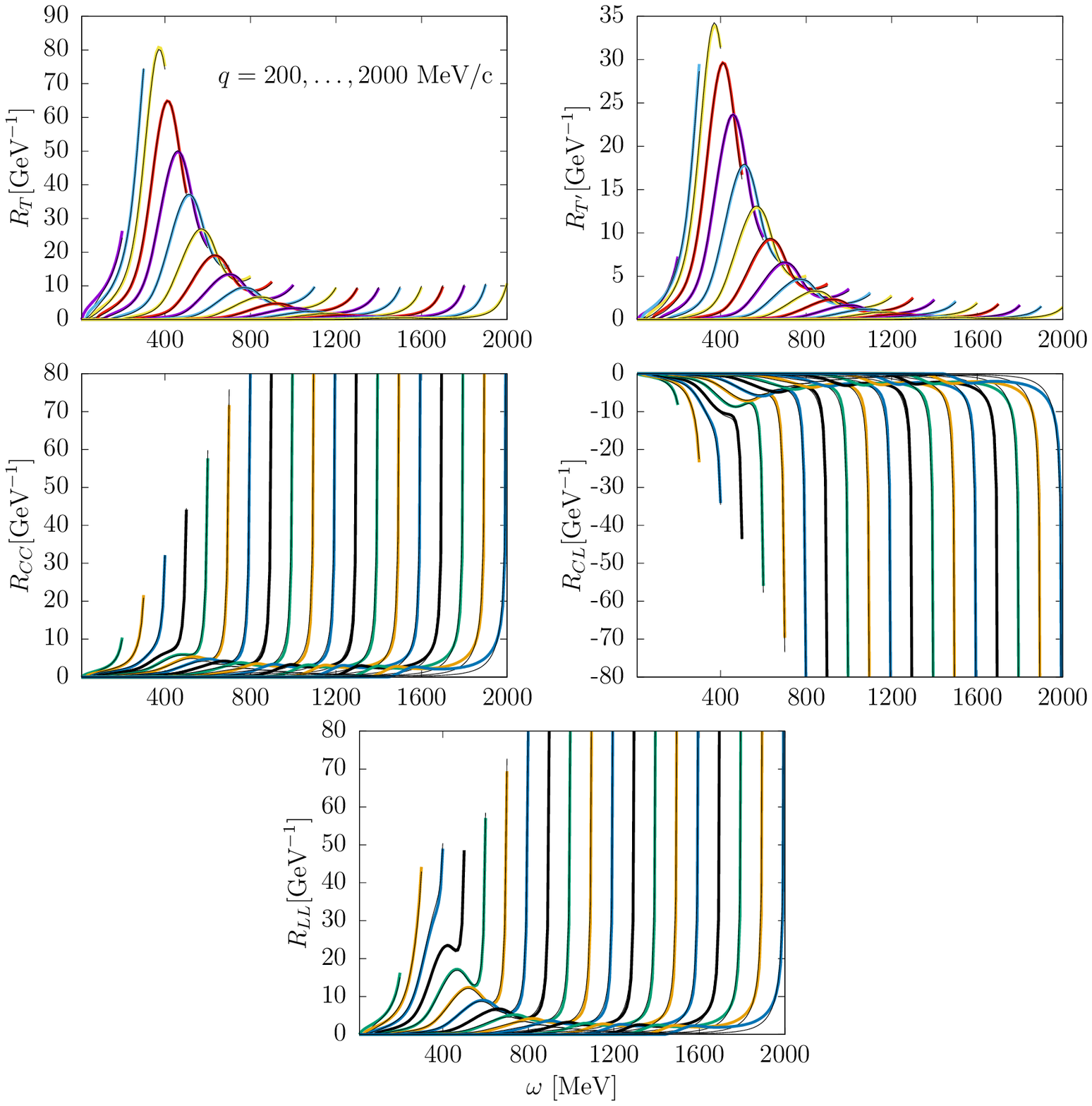}
\caption{ Comparison of 
the five 2p2h response functions for CC neutrino scattering
  from $^{12}$C, computed with the RMF model and with the
  semi-empirical formula. They are plotted as a function of $\omega$
  for fixed values of thew momentum transfer $q=200,\ldots,2000$ MeV/c.}
\label{revalida3}
\end{figure*}

\begin{figure*}
\includegraphics[width=7cm, bb=145 440 393 775]{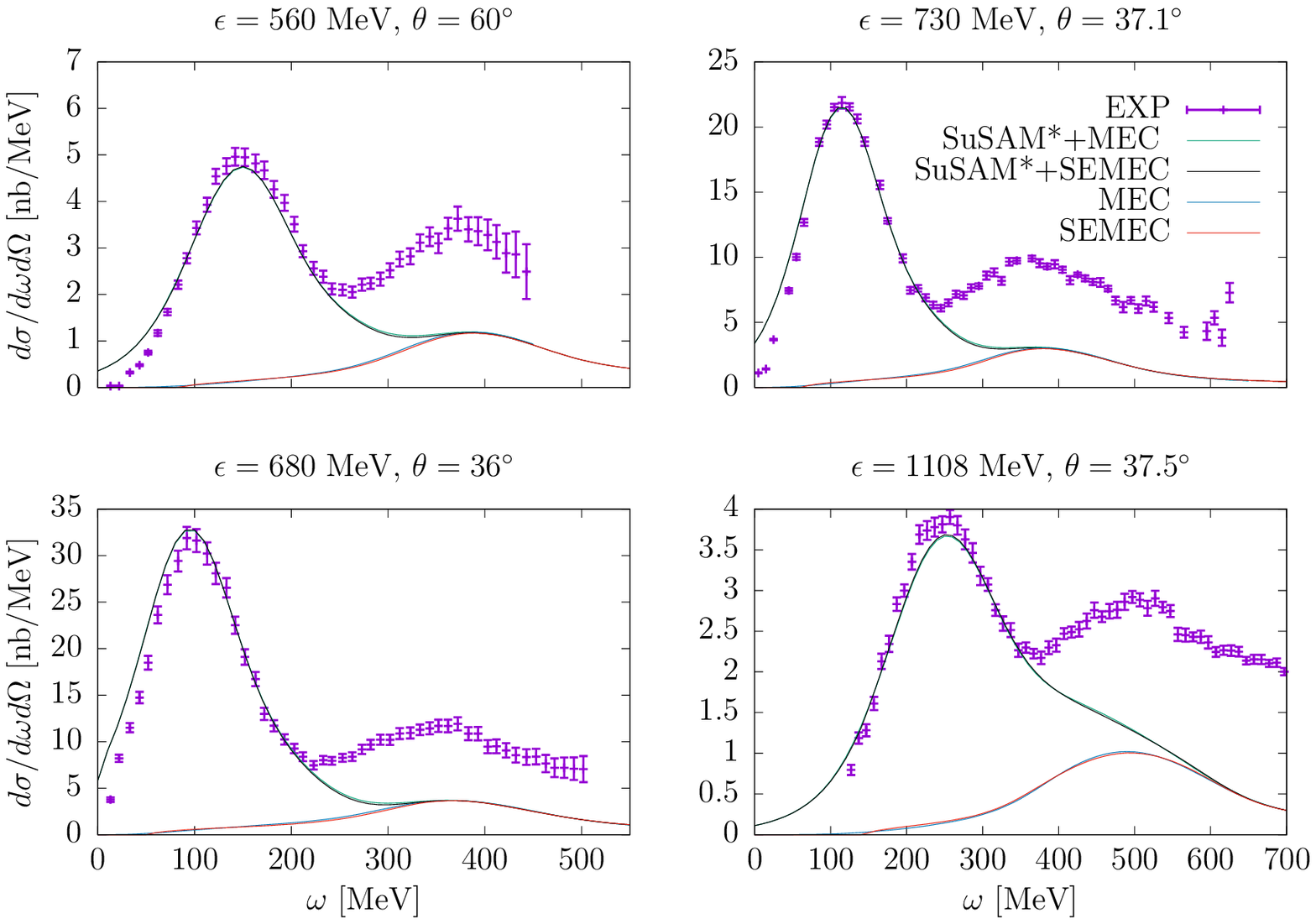}
\caption{ 
Cross section for $(e, e ')$ scattering off $^{12}$C for various
  kinematics as a function of omega. The quasi-elastic contribution
  with emission of a particle is shown using the SuSAM* model of
  reference \cite{Mar21} and the contribution of channel 2p2h, calculated both
  with the exact RMF (MEC), and with the semi-empirical formula
  (SE-MEC). Experimental data are from \cite{archive,archive2,Ben08}. 
}
\label{revalida4}
\end{figure*}

\begin{figure*}
\includegraphics[width=7cm, bb=145 440 393 775]{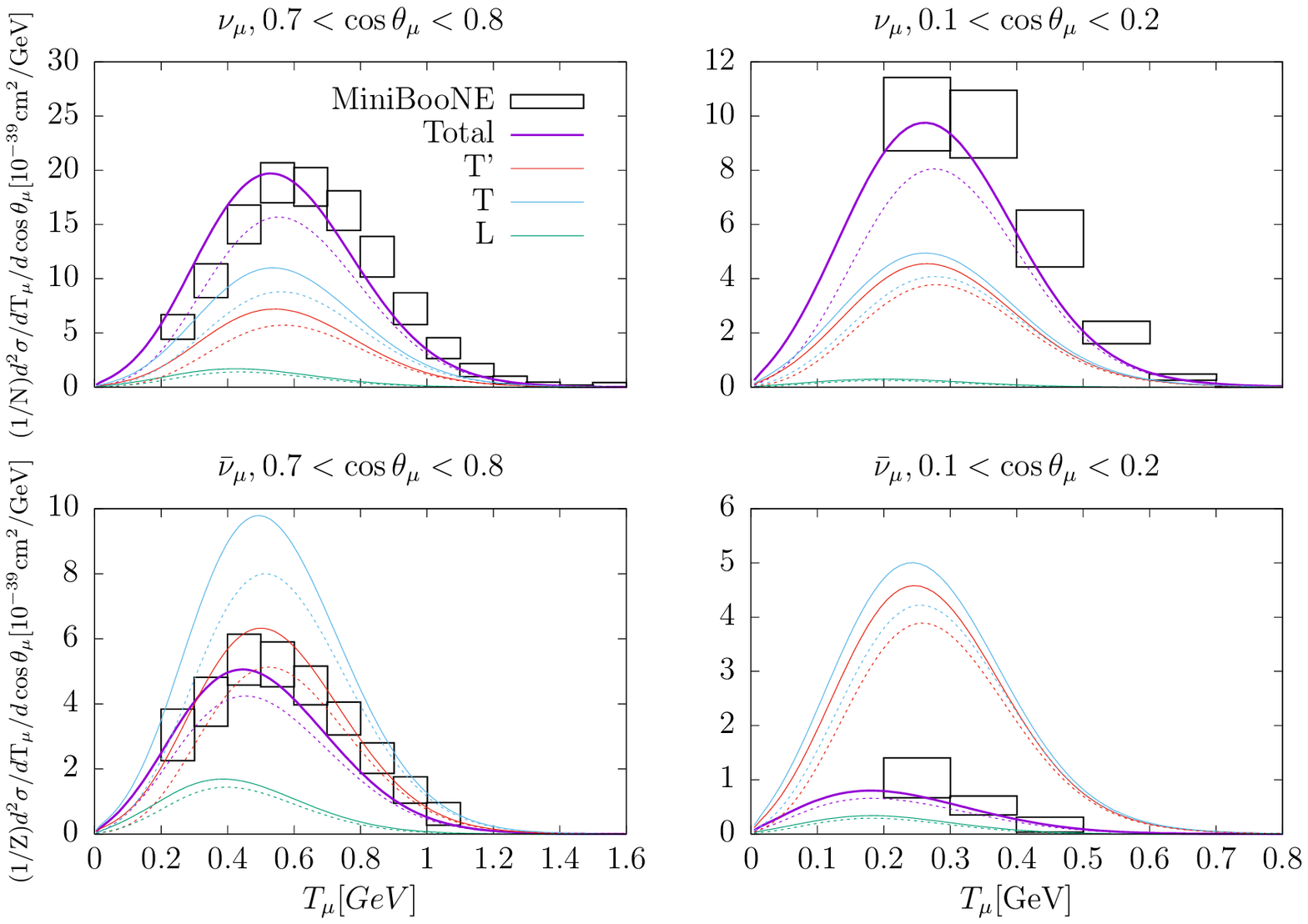}
\caption{ Quasielastic neutrino and antineutrino, differential cross
  section integrated over the neutrino flux of the MiniBooNE
  experiment for selected kinematics of the scattering angle bins.  we
  show the separate $T$, $T'$ and longitudinal responses (L = CC + CL
  + LL) with and without MEC.
  The curves without MEC (dotted lines) have been computed using the SuSAM* model of ref. \cite{Mar21}. The curves with MEC  (solid lines)
  have been calculated with
  the SE formula of the 2p2h responses.
Experimental data from refs. \cite{Agu10,Agu13}.
 }
\label{revalida5}
\end{figure*}

\begin{figure*}
\includegraphics[width=8cm, bb=120 270 430 780]{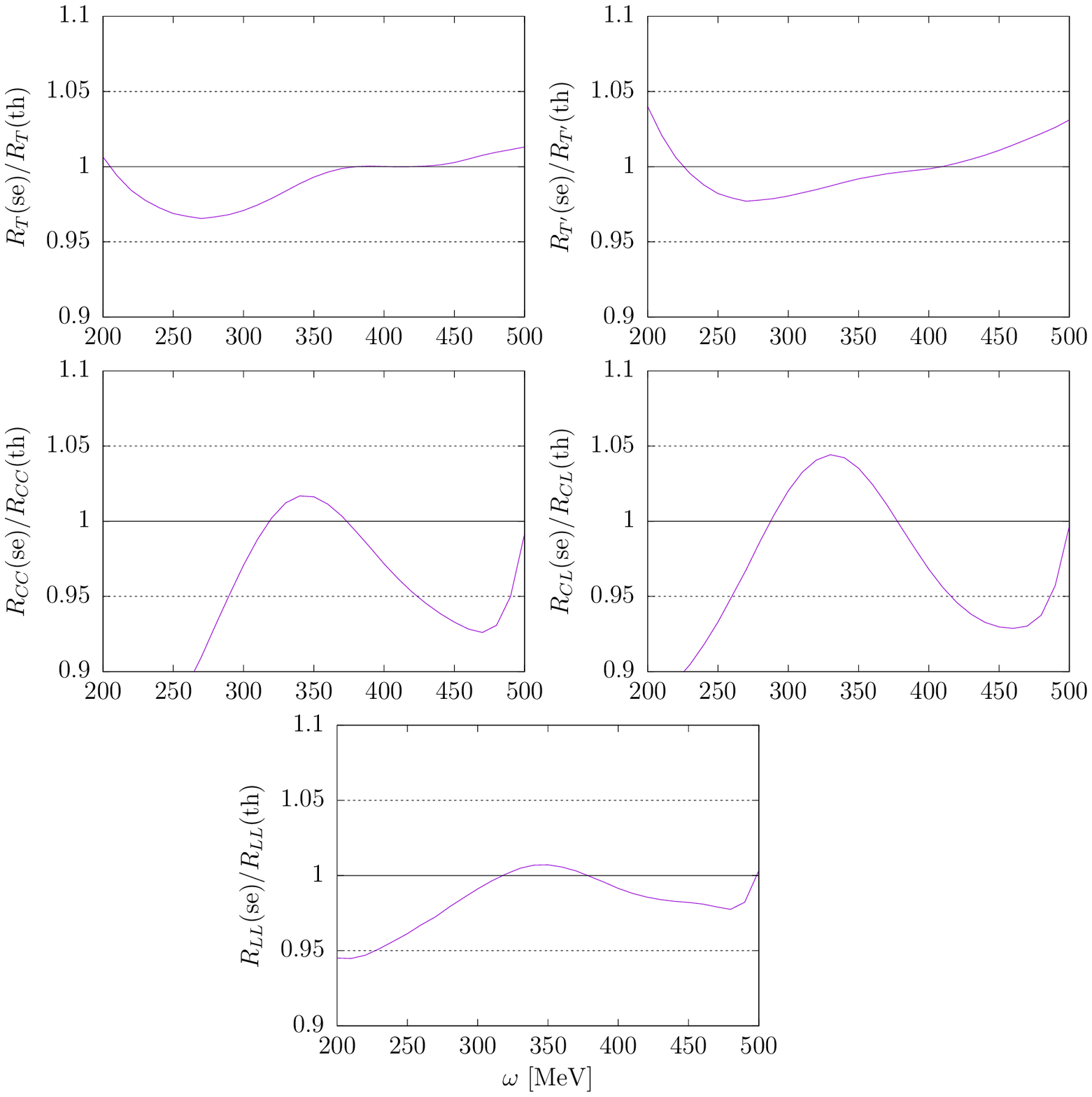}\\
\includegraphics[width=8cm, bb=120 580 430 780]{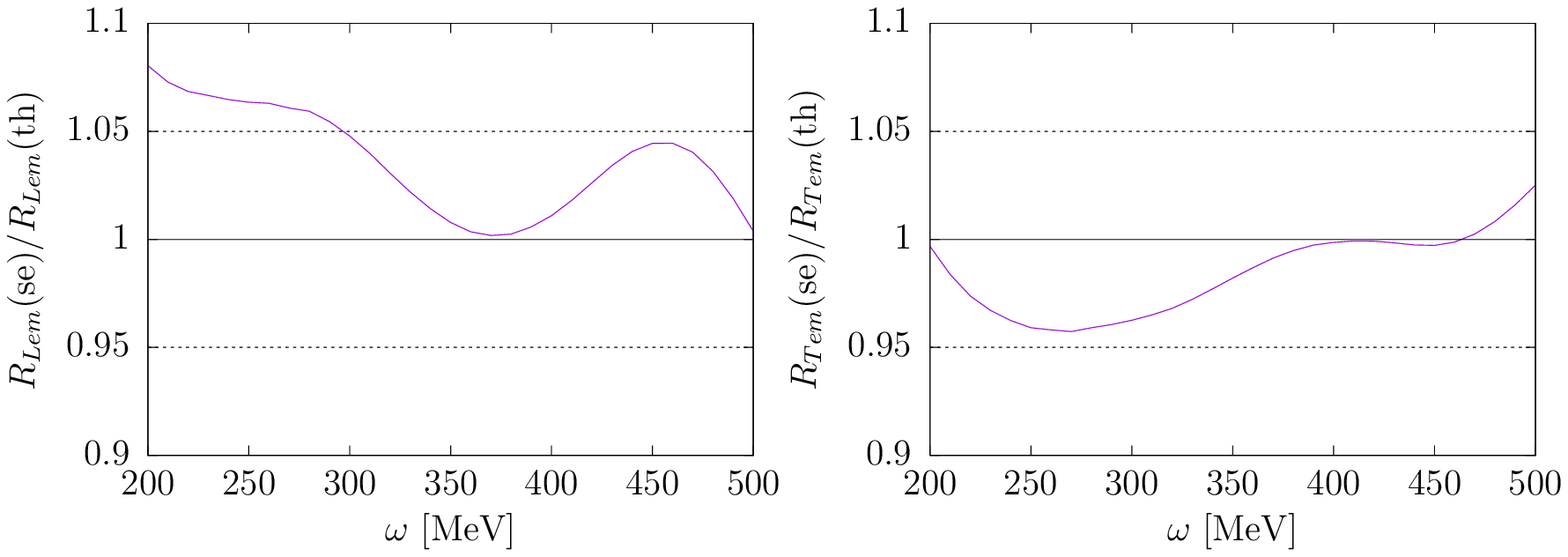}
\caption{ Quotient between the semi-empirical (se) and theoretical
  (th) 2p2h response functions for CC neutrino and electron
  scattering. For $q=500$ MeV/c as a function of $\omega$.
 }
\label{revalida6}
\end{figure*}

In this section we obtain the values of the coefficients of the semi
empirical formulas (\ref{rtforv}--\ref{rccbackspa}),
and the parameters $\Gamma$, and $\Sigma$ in the averaged $\Delta$ propagator
(\ref{averaged_denom},\ref{afrozen},\ref{bfrozen}).
  Since the
experimental responses are not available, nor are they possible to
obtain phenomenologically, then the coefficients cannot be obtained
directly from data. Therefore the only possibility is to fit a
theoretical model. In our case, the interest is to obtain a
parameterization of the responses to make theoretical predictions of
neutrino cross sections in a computationally fast way.
All numerical results in this section correspond to $^{12}$C with $k_F = 225$
MeV/c, $M^* = 0.8$ and $E_v = 141$ MeV, fitted in \cite{Mar21}.

First we have calculated the ``exact'' sub-responses for a set of
kinematics, $(q,\omega)$, performing numerical integrations in seven
dimensions in the RMF of nuclear matter described in Sect. II. Our
computer code calculates the seven total 2p2h responses of neutrino CC
scattering, $R_{K}$, $K=CC, CL, LL, T, T'$, and electrons, $R^L_{\rm
  em}$ and $R^T_{\rm em}$, using Eq. (\ref{responses}).  Our numerical
code allows to include in the calculation some specific Feynman
diagrams and exclude others.  To compute the sub-responses, we have
performed three runs with the individual currents, SP, $\Delta F$ and
$\Delta B$, and three more runs with the pairs of currents (SP +
$\Delta F$), (SP + $\Delta B$), and ($\Delta F$ + $\Delta B$).
Subtracting the separate contributions of the single currents, the
interference sub-responses are obtained. For instance
\begin{equation}
R_{SP-\Delta F} = R_{SP + \Delta F} - R_{SP} - R_{\Delta F}    
\end{equation}
An additional run is performed with the full MEC to get the complete
result.  Each run requires to compute the seven response functions in
a grid of $(q,\omega)$ values, with $q=200,\ldots,2000$ MeV/c in steps
of $\Delta q =100$ and $\omega=10,\ldots,q$ in steps of
$\Delta\omega=10$ MeV.  The grid contains about 15000 kinematical
points $(q_i,\omega_i)$.  Note that the computation for each
kinematical point takes an average of 5 minutes on our
high-performance processors.  This means each run requires about 52
days in one processor, and one year for the seven runs. We have used
the PROTEUS scientific computing cloud of the ic1 \cite{proteus} (2300
processors with total 90000 GFLOPs), allowing to do the calculation in a
few days.

Once we have stored the tables of the exact sub-responses in the grid,
we fit the coefficients of the semi-empirical formula for fixed
$q$. The fit is made by minimizing a $\chi^2$ function for each
subresponse separately, thus obtaining the coefficients
$\tilde{C}_i(q)$.  The $R^{T,VV}_{\Delta F}$ responses are used to fit
the effective width, $\Gamma_V(q)$ and shift, $\Sigma_V(q)$ of the
averaged delta propagator as well. The same procedure is followed to
fit the $AA$ and $VA$ effective widths and shifts using the responses
$R^{T,AA}_{\Delta F}$, and $R^{T',VA}_{\Delta F}$, respectively.
Therefore, in the transverse sub-responses $\Delta F$, four
coefficients are being adjusted simultaneously.  Once the widths and
shifts have been fitted in this way, they are set to that value in all
the sub-responses.

The coefficients obtained in the fit are tabulated in Appendix
\ref{apendice1}.  In table \ref{t0} the effective width, $\Gamma(q)$,
and shift, $\Sigma(q)$, of the averaged $\Delta$ propagator, $G_{\rm
  av}(q,\omega)$, are given for $VV$, $AA$ or $VA$ responses.  The
coefficients $\tilde{C}_i$ are provided in tables \ref{t1} (for the
$R_T^{VV}$ response function), \ref{t2} (for $R_T^{AA}$), \ref{t3}
(for $R_{CC}^{VV}$), \ref{t4} (for $R_{CC}^{AA}$), \ref{t5} (for
$R_{CL}^{AA}$), \ref{t6} (for $R_{LL}^{AA}$), and \ref{t7} (for
$R_{T'}^{VA}$). The tables are given as a function of $q$ and can be
interpolated for other $q$-values. We also provide polynomial
parametrizations in Appendix \ref{apendice2}.

In Fig. \ref{revalida1} we plot the coefficients of the semi empirical
formula for the case of the sub-responses $R^T$ for $VV$ and $AA$
cases. We plot the coefficients $\tilde{C}_i$, and the effective
widths, $\Gamma_V$, $\Gamma_A$, and shifts, $\Sigma_V$ , $\Sigma_A$,
of the averaged $\Delta$ propagator as a function of the momentum
transfer $q$.  Smooth dependence on $q$ is observed, except for very
low $q$ below 300 MeV/c where the dependence is more abrupt in some
cases, specifically in the case of the widths and shifts of the
$\Delta$ propagator for $q <300$ MeV/c.  The reason why there is an abrupt
change for low momentum in the $\Delta$ current coefficients, is
because the peak of the delta is not reached below $q = 300$ MeV/c, since at
least the transferred energy must be large enough to produce the
$\Delta$. In the present RMF model, this is 
$\omega = m_\Delta-m_N^*-E_v \simeq 340$ MeV.  
Therefore these coefficients are
less restricted and their value has greater indeterminacy when making
the fit.
Similar results ---not shown in Fig. \ref{revalida1}--- 
are obtained for the dependence on $q$ 
of the rest of the coefficients of the semi-empirical formula.

The most relevant coefficients for calculating the responses are those
corresponding to the $\Delta$-forward transverse responses ($T$ and $T'$),
because these subresponses are dominant in the $\Delta$ region.  These
coefficients are $\tilde{C}_{1 Vi}$, $\tilde{C}_{1 Ai}$, and
$\tilde{C}_{1 VAi}$.  They increase moderately with $q$ and their
values vary between $\tilde{C}_i\simeq 0$ and 100 for $200 \leq q \leq 2000$
MeV/c.

In Figures \ref{revalida2a} and \ref{revalida2b} we show, as an
example, all the responses and sub-responses as a function of $\omega$
for two values of the momentum transfer $q=500$ and 1000 MeV/c,
respectively.  The dominant sub-responses in the responses $T$ and
$T'$ are the delta-forward ones, while the axial delta-forward and
seagull-pionic dominate in the longitudinal responses $CC$, $CL$ and
$LL$.  The rest of the sub-responses give a very small contribution to
the total, and could in principle be neglected, although we have
included all in the calculation.  In the figures we plot the
the exact result  and semi-empirical formula, 
using
the parametrization of the $\tilde{C}_i$ the Appendix \ref{apendice2},
for each one of the
dominant sub-responses and also for the total responses. 

In Fig. \ref{revalida3} it is seen that the total responses are well
described by the semi-empirical formula in the range of
$q=200,\ldots,2000$ MeV/c considered in the present work.  The five
2p2h response functions for CC neutrino scattering from $^{12}$C,
computed with the RMF model and with the semi-empirical formula are
mostly identical in the scale of the figure.  This indicates that the
semi-empirical formula can be used with guarantees to calculate the
cross section in the 2p2h channel.  For other values of $q$ it is
enough to interpolate the tables of the coefficients of Appendix
\ref{apendice1}, or to use the polynomial parametrizations of Appendix
\ref{apendice2}. In figure 6 we give an example of how the formula
works in the case of $(e,e')$ cross section of $^{12}$C for various
kinematics.  The electron energy and the scattering angle are fixed in
the experiment.  When changing omega, the momentum transfer is not
constant, but depends on the three variables continuously.  Then it is
necessary to interpolate the coefficients of the semi-empirical
formula to calculate the 2p2h cross section. In the figure we used
polynomial interpolation.

The greatest utility of the SE formula is to calculate the 2p2h
interaction with neutrinos, since the neutrino flux implies an
integration on the incident energy. The integrated cross section in
the flux is shown in figure \ref{revalida5}. There we show the
transverse and longitudinal contributions to the cross section with
and without 2p2h MEC.  The effect of MEC responses, computed with the
SE formula, is to increase the cross section of about $\sim 20 \%$,
depending on the kinematics.
Note that these results have been obtained using $C^5_A (0) = 1.2$ for
the axial $\Delta$ coupling. But in ref. \cite{Hernandez:2007qq}
it was found that a value of 0.89 was more adequate according to the pion
emission data. If this value is used with the semi-empirical formula,
the effect of the axial MEC contribution  in Fig. \ref{revalida5} 
would be reduced by almost one half.
Note that the L-contribution of the MEC is very small and almost
negligible and could be omitted in the calculation.

To visualize the quality of the semi-empirical formulas, we show in
figure \ref{revalida6} the quotient between the semi-empirical formula
(se) and the exact result (th), for $q=500$ MeV/c as a function of
$\omega$.  In the zones dominated by the transverse responses $T$ and
$T'$ at the peak of the delta, the quotient is very close to one.  For
the $\omega$ values where the responses are appreciable, the quotient
is practically one, because the coefficients have been adjusted.  The
quotient deviates from one only for $\omega$-values where the responses
are not important or negligible.

\begin{figure*}
\includegraphics[width=11cm, bb=40 400 530 780]{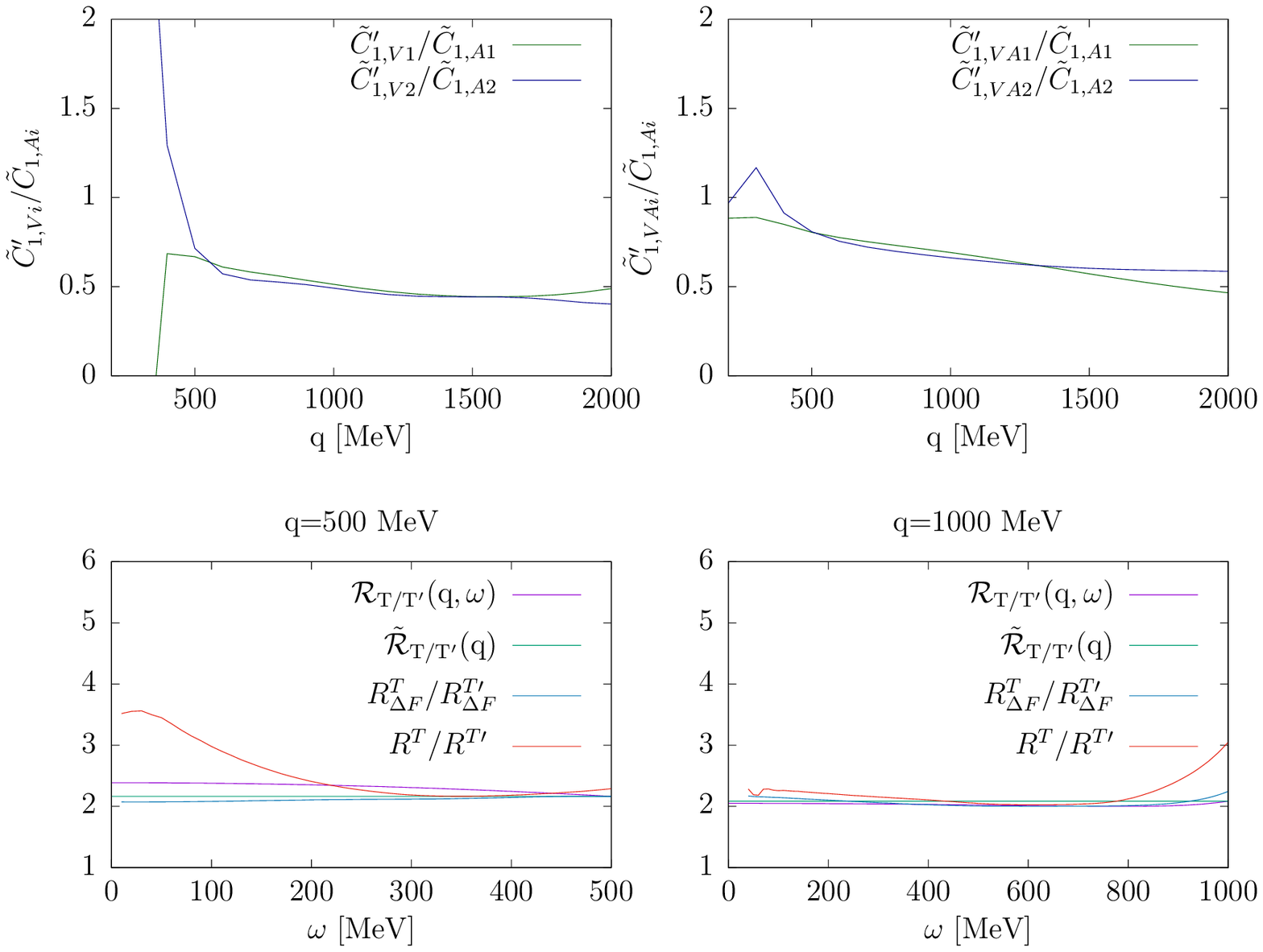}
\caption{Top: Relation between the coefficients of the $\Delta F$
  subresponses $R^{TVV}_{\Delta F}$ and $R^{T'VA}_{\Delta F}$, with
  respect to $R^{TAA}_{\Delta F}$.  Bottom: Comparison of the quotient
  $R^T/R^{T'}$ with several approximations (see text). The ${\cal
    R}_{T/T'}$ and $\tilde{\cal R}_{T/T'}$ functions have been obtained
  from the semi-empirical formulas of the $\Delta F$ sub-responses. }
\label{revalida8a}
\end{figure*}

The semi-empirical formula also allows studying the relative behavior
between the different contributions or sub-responses.  In particular
one can find relations between the dominant sub-responses
$R^{TVV}_{\Delta F}$, $R^{T'VA}_{\Delta F}$, and
 $R^{TAA}_{\Delta F}$ as follows. First note that
the only difference between the vector and axial parts of the $\Delta F$
 current, 
Eq. (\ref{deltaF}), 
is in the electroweak vertex (\ref{gamma-delta}), 
$\Gamma^{\beta\mu}(Q)=\Gamma^{\beta\mu}_V(Q)+\Gamma^{\beta\mu}_A(Q)$, with
\begin{equation}
\Gamma^{\beta\mu}_V(Q)=
\frac{C^V_3}{m_N}
\left(g^{\beta\mu}\Qbar-Q^\beta\gamma^\mu\right)\gamma_5, 
\kern 1cm
\Gamma^{\beta\mu}_A(Q)=
C^A_5 g^{\beta\mu},
\end{equation}
Then the vector current $\Delta F$ is expected to behave roughly like
$q/m$ with respect to the axial current.  As a consequence the
coefficients of the semi-empirical $R^{TVV}_{\Delta F}$ and $R^{T'VA}_{\Delta F}$
responses would contain a factor $(q / m) ^ 2$ and $q/m$,
respectively, with respect to the $R^{TAA}_{\Delta F}$ coefficients.
Therefore, if we define the coefficients
\begin{equation}
\tilde{C}'_{1,Vi} \equiv \frac{\tilde{C}_{1,Vi}}{ (q/m_N)^2},
\kern 1cm
\tilde{C}'_{1,VAi} \equiv \frac{\tilde{C}_{1,VAi}}{ (q/m_N)},
\end{equation}
then one expect the quotient
$\tilde{C}'_{1,Vi}/\tilde{C}_{1,Ai}$, and
$\tilde{C}'_{1,VAi}/\tilde{C}_{1,Ai} $ to be approximately independent of $q$.
This is shown in figure \ref{revalida8a},
 where we plot these quotients, 
and see that 
they  are roughly
\begin{eqnarray}
\frac{\tilde{C}'_{1,Vi}}{\tilde{C}_{1,Ai}} &\simeq& \frac12
\label{c1v} \\
\frac{\tilde{C}'_{1,VAi}}{\tilde{C}_{1,Ai}}
&\simeq&
 \frac{1}{\sqrt{2}}
\label{c1va}
\end{eqnarray}

\begin{figure*}
\includegraphics[width=6cm, bb=180 600 420 770]{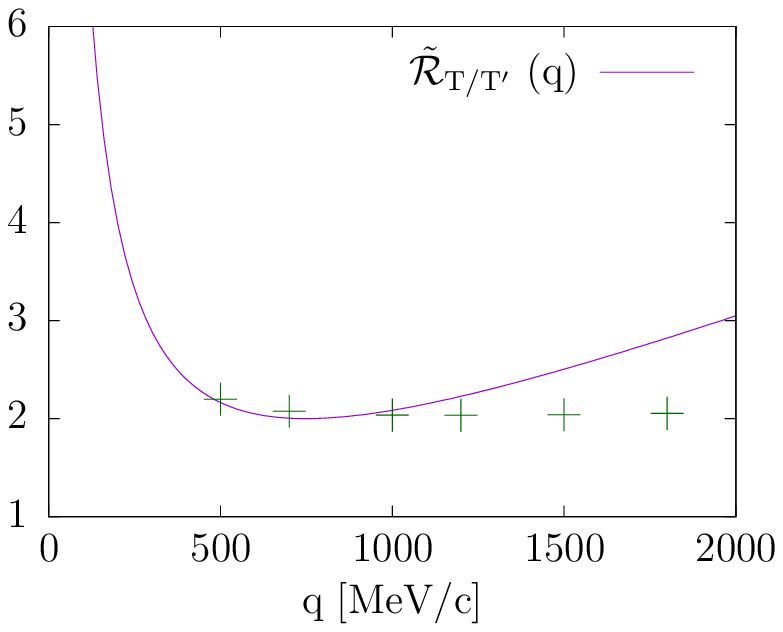}
\caption{ The semi-empirical approximation $\tilde{\cal R}_{T/T'}$ of the
  quotient between the $T$ and $T'$ responses, as a function of $q$.  }
\label{revalida8c}
\end{figure*}

If we neglect the small contributions of the $SP$ and $\Delta$B
diagrams, the semi-empirical formulas
(\ref{rtforv},\ref{rtfora},\ref{rtpfor}) allow to make quantitative 
estimations of relation between the $T$ and $T'$ responses.
In fact the following approximate formulas can be
obtained between $R^{T,VV}$ and  $R^{T,AA}$ 
\begin{equation} \label{vvaa}
\frac{R^{T,AA}}{R^{T,VV}}
\simeq
\frac{R^{T,AA}_{\Delta F}}{R^{T,VV}_{\Delta F}}
\simeq
\frac{(C_5^A)^2}{ (C_3^V)^2\frac12 \left(\frac{q}{m_N}\right)^2}
\end{equation}
and also the following relation between the $T$ and $T'$ responses.
\begin{equation} \label{calr}
\frac{R^T}{R^{T'}}
\simeq
\frac{R^T_{\Delta F}}{R^{T'}_{\Delta F}}
\simeq
\frac{ (C_3^V)^2\frac12 \left(\frac{q}{m_N}\right)^2+(C_5^A)^2}
{  C_3^V C_5^A \frac1{\sqrt{2}} \left(\frac{q}{m_N}\right)}
\equiv {\cal R}_{T/T'}(q,\omega)
\end{equation}
Where we have defined he function ${\cal R}_{T/T'}(q,\omega)$ that
represents the approximate quotient between the sub-responses $T$ and
$T'$ for the $\Delta F$ diagrams.  To obtain Eq. (\ref{calr}) we have
used the empirical relations (\ref{c1v},\ref{c1va}) between the
corresponding coefficients, and we have assumed that the averaged
$\Delta$ propagators are similar for the $VV$, $AA$ and $VA$
responses, and they cancel out in the numerator and denominator.  The
function ${\cal R}$ depends on $\omega$ through the $Q^2$ dependence
of the $\Delta$ form factors $C_3^V(Q^2)$ and $C_5^A(Q^2)$. The
comparison between this relationship and the exact result is also
shown in Fig. \ref{revalida8a} for two values of the momentum
transfer.

If we also assume that the form factors have an approximately similar
dependence on $Q^2$, and that this dependence 
is canceled in the numerator and the
denominator, we can simply use the values at the origin, $Q^2=0$, of the form
factors 
to obtain a simple approximate formula for the relationship between
T and T' responses.  
In fact, inserting the values
$C_5^A(0)=1.2$, and $C_3^V(0)=2.13$ 
in Eq. (\ref{calr}), 
we can write 
\begin{equation} \label{cociente}
\frac{R^T}{R^{T'}}
\simeq
\frac{ (2.13)^2\frac12 \left(\frac{q}{m_N}\right)^2+(1.2)^2}
{  2.13 \times 1.2 \frac1{\sqrt{2}} \left(\frac{q}{m_N}\right)}
\equiv \tilde{\cal R}_{T/T'}(q)
\end{equation}
Note that this approximation does not depend on $\omega$, and this
relation is compared in Fig. \ref{revalida8c} with the exact result
computed for $\omega$ at the maximum of $\Delta$ peak.  We see that
the formula (\ref{cociente}) is valid for $q$ between $500$ and $1100$
MeV/c, where $\tilde{\cal R}_{T/T'}\simeq 2$. 
Note that this value depends on the value of $C^5_A$, and here we have used 1.2.
For larger $q$ the approximations fail because in this region
the dominance of the $\Delta F$ starts to decay.

Note that equations (\ref{calr},\ref{cociente}) relate $R^T$ and
$R^{T'}$, similarly to eqs. (8,9) of ref. \cite{Gal16}, but with the
$\Delta$ form factors instead of the nucleon form factors, and with
similar kinematic factors. In our case we see, from figs. 9,10, that
this relationship is approximately $ R^T \simeq 2 R^{T'}$ in the
region of the $\Delta$ peak.  This opens the way to determine the 2p2h
response of neutrinos from fits of the corresponding response
in the electromagnetic channel.

\begin{figure*}
\includegraphics[width=14cm, bb=31 430 534 780]{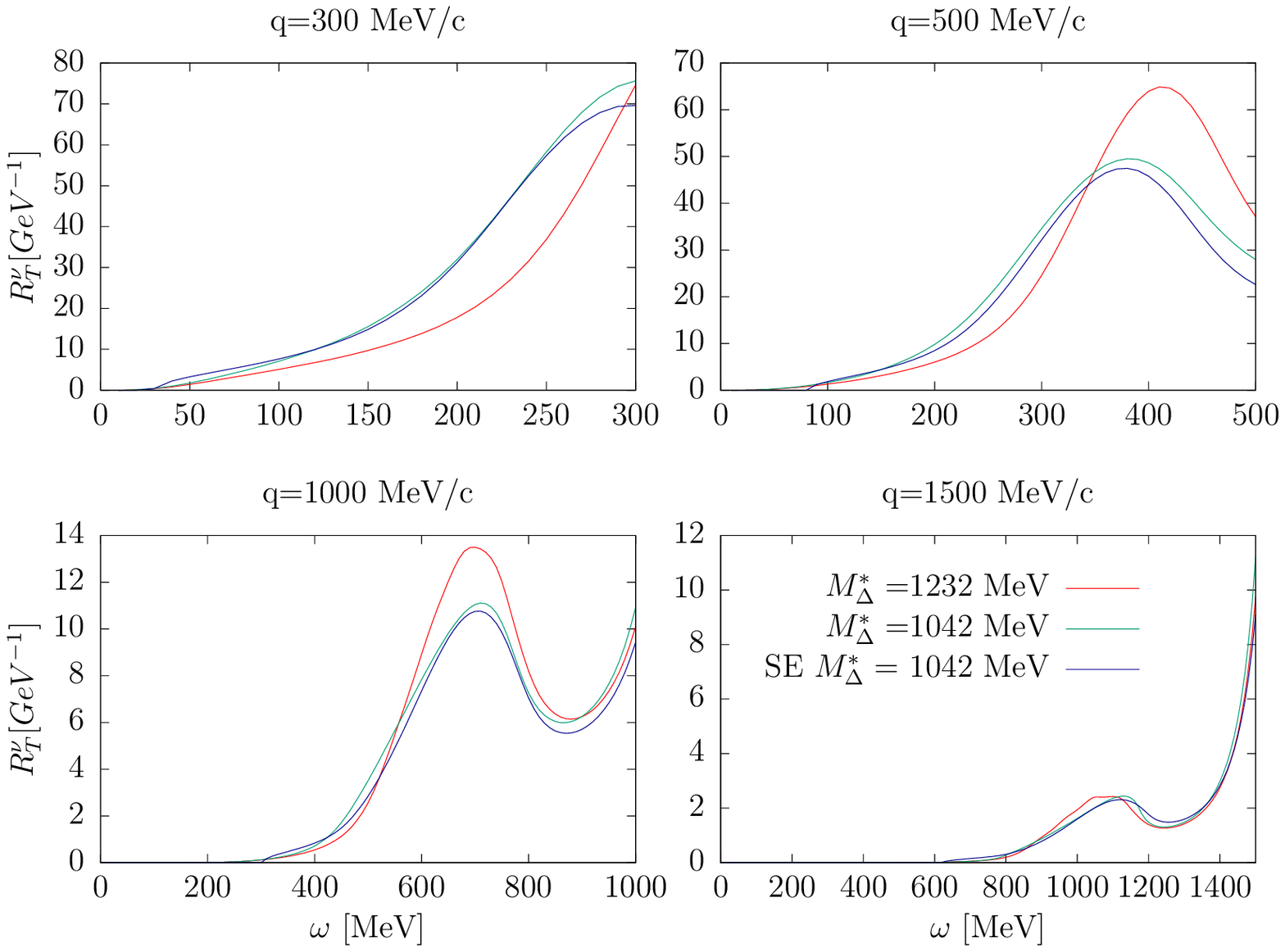}
\caption{ Comparison of the transverse response functions including or
  not the $\Delta$ self-energy in the MEC, in the case of universal
  coupling, $E_v^\Delta=E_v$, and $M_\Delta^*=1042$ MeV.  The results
  of the modified semiempirical formula for interacting $\Delta$ are
  also shown. }
\label{revalida9}
\end{figure*}

Finally, in this work we have considered the non-interacting free
$\Delta$ in the MEC. The semi-empirical formula can be extended to
include the case of a $\Delta$ interacting with the mean field
\cite{Weh93,Kim96}.  In this case the $\Delta$ acquires effective
mass, $M_\Delta^*$, and vector energy, $E_v^\Delta$ (see appendix
\ref{apendice0}). In the simplest case of universal coupling, the
$\Delta$ scalar and vector energies are the same as those of the
nucleon.  In Fig. \ref{revalida9} we show effect of including the
$\Delta$ self-energy for universal coupling, compared to the case of
free $\Delta$.  In fig. \ref{revalida9} we also compare with the
extended semi-empirical formula of appendix \ref{apendice0}.  The
effect of the $\Delta$ interaction is a $q$-dependent shift, from -50 MeV,
for small $q$, up to +50, for large $q$. 
The height of the transverse response
decreases for low $q$ by $\simeq 30\%$. The semi-empirical formula allows
to easily study the dependence of the  responses for other values of
the $\Delta$ interaction with the mean field.

\section{Conclusions}

In this article we have proposed a semi-empirical formula to
approximately calculate the 2p2h responses in neutrino scattering. The
formula is based on classifying the contribution of Feynman diagrams
of MEC with a similar structure, in terms of the same electroweak form
factor and the same number of delta-forward propagators, obtaining in
our case six contributions, or sub-responses, from our three types of
diagrams: delta-forward, seagull-pionic, and delta-backward.

It is proposed that each sub-response is the product of the phase
space function of two-particles, electroweak form factors and averaged
delta propagators, multiplied by coefficients that only depend on q.
These coefficients are fitted with a relativistic mean field model of
nuclear matter for the 2p2h responses.  In the SE formula we use an
approximation of the phase space (frozen nucleon approximation) and a
model for the average propagator of the delta forward.

The hypothesis that the coefficients do not depend on $\omega$ is
generally fulfilled, with two minor exceptions in the transverse SP
sub-responses, where a corrective term with a second degree polynomial
in $\omega$ is added.  Altogether we have shown that the dependence on
$\omega$ of the 2p2h responses comes mainly from these three elements:
phase space, form factor and averaged propagator of the delta forward,
and that's it.

Having extracted in the sub-responses everything that can be factorized,
the semi-empirical formulas explicitly contain the dependence 
on the Fermi momentum,
the number of particles, the effective mass and vector energy of the
RMF, the electroweak form factors, and the coupling constants, 
in addition to the explicit 
dependence on $q$ and $\omega$ through the phase space and the averaged $\Delta$
propagator.

The semiempirical formula has been obtained for free $\Delta$, but then
we have generalized it to include a $\Delta$ interacting with the mean
field, which already depends on the $\Delta$ effective mass and
its vector energy.

An advantage of the semi-empirical formula is that it provides the
2p2h responses in a model in an easily reproducible way and that it
also allows many of the model's parameters to be varied at will.

The semi-empirical formula could be extended by including more
diagrams or contributions to the MEC, for example $\rho$-meson exchange,
correlation diagrams, excitation of other nucleon resonances, simply by adding
the corresponding form factors, couplings and propagators to construct
the sub-responses of the new contributions.
The SE formula could also be fitted with other models to recalculate 
the coefficients.

The SE formulas are promising for application in neutrino calculations and
event simulators
 because they are analytical and allow modifying 
physical parameters of the model.
We are currently studying other nuclei and we have checked that the
formula is valid in symmetric nuclei, $N=Z$, with the same coefficients
of this article, and only the averaged $\Delta$ propagator must be modified
by readjusting the effective width and shift.

In the future we will apply the semi-empirical formula 
to the case of asymmetric matter, $N\neq Z$, which is of interest
for current and future neutrino experiments, 
since in this case the emission channels of
pp, pn and nn have to be treated separately.

\section{Acknowledgments}

This work has been supported by the Spanish Agencia Estatal de investigacion
(D.O.I. 10.13039/501100011033, 
Grants Nos. FIS2017-85053-C2-1-P and PID2020-114767GB-I00), 
the Junta de Andalucia (Grant No. FQM-225),
and the European Regional Development Funds (Grant No. A-FQM-390-UGR20).
V.L.M.-C. acknowledges a contract funded by Agencia Estatal
de Investigacion and European Social Fund.

\appendix

\section{Semiempirical formula for $\Delta$ in the medium}

\label{apendice0}

In this appendix we explain how the semi-empirical formula must be
modified to include the effective mass of the $\Delta$ and its vector
energy.  In this work we have neglected the interaction of the
$\Delta$, treating it as a free particle. If we assume that the
$\Delta$ interacts with the scalar and vector fields of the
relativistic mean field, the $\Delta$ acquires an effective mass and a
vector energy \cite{Weh93,Kim96}
\begin{eqnarray}
M_\Delta^*  &=& M_\Delta - g_s^\Delta\phi_0  \\
E_{RMF}^\Delta  &=& E^\Delta + E_v^\Delta
\end{eqnarray}
with $E^\Delta= \sqrt{ p_\Delta^2+M_\Delta^{*2}}$ is the on-shell
energy of the intermediate $\Delta$ isobar with momentum $\np_\Delta$.

In this case, the delta current would be modified by substituting the
propagator of the free delta , Eq. (\ref{delta_prop}), for the
propagator in the medium, which is the following \cite{Weh93}
\begin{equation}\label{delta_prop2}
 G_{\alpha\beta}(P)= \frac{{\cal P}_{\alpha\beta}(P^*)}{P^{*2}-
 M^{*2}_\Delta+i M^*_\Delta \Gamma_\Delta(P^{*2})+
 \frac{\Gamma_{\Delta}(P^{*2})^2}{4}} \, .
\end{equation}
The projector  ${\cal P}_{\alpha\beta}(P^*)$ 
is now
\begin{eqnarray}
{\cal P}_{\alpha\beta}(P^*)&=&-(\Pbar^*+M^*_\Delta)
\left[g_{\alpha\beta}-\frac13\gamma_\alpha\gamma_\beta-
\frac23\frac{P^*_\alpha P^*_\beta}{M^{*2}_\Delta}\right.
+\left.
\frac13\frac{P^*_\alpha\gamma_\beta-
P^*_\beta\gamma_\alpha}{M^*_\Delta}\right].
\end{eqnarray}
where $P^{*\mu}= P^\mu-\delta_{\mu,0}E_v^\Delta$.    With the $\Delta$ in the
medium the peak position of the $\Delta$-forward response is expected
at
\begin{equation}
\omega= \sqrt{q^2+M_\Delta^{*2}}-m_N^*-E_v+E_v^\Delta
\end{equation}
In the particular case of a free $\Delta$ we recover the original position
of the peak at
\begin{equation}
\omega= \sqrt{q^2+M_\Delta^{2}}-m_N^*-E_v
\end{equation}
We see that the position of the $\Delta$-forward peak depends on $q$
and the values of the vector energies and the effective masses. 
In this case the averaged $\Delta$ propagator,
Eq. (\ref{averaged_denom}), is still calculated with the same formula,
but changing the values of the parameters $a$ and $b$ to include the
vector energy and the effective mass of the $\Delta$ in the mean
field
\begin{eqnarray}
\kern -0.8cm 
a &\equiv& 
m^{*2}_N+(\omega+E_v-E_v^\Delta)^2-q^2+2m^*_N(\omega+E_v-E_v^\Delta+\Sigma)
-M^{*2}_\Delta+
\frac{\Gamma^2}{4}
\label{afrozen2}\\
\kern -0.8cm 
b &\equiv& M^*_\Delta \Gamma
\label{bfrozen2} \,.
\end{eqnarray}

Note that the $\Delta$ effective mass also appears in the numerator of
the $\Delta$ propagator, and this modifies the values of the matrix
elements of the $\Delta$ forward, and the values of the responses.  We
have computed the new responses.  For simplicity we have assumed
universal coupling, where the scalar and vector selfenergies of the
$\Delta$ are the same as the nucleon, and also the case where only the
vector energies are equal.  For universal coupling
$E_v^\Delta=E_v=141$ MeV, and $M_\Delta^*=1042$ MeV.  We have found
that the the semiempirical formula can be easily modified to include
these cases in the following lines, by replacing the parameters of the
$\Delta$-forward responses
\begin{eqnarray}
\tilde{C}_{1,ij} 
&\rightarrow&
 \left( \frac{M_\Delta^*}{M_\Delta} \right)^2
\tilde{C}_{1,ij}  
\\
\tilde{C}_{4,ij} 
&\rightarrow&
 \left(\frac{M_\Delta^*}{M_\Delta}\right)
\tilde{C}_{4,ij}  
\\
\tilde{C}_{5,ij} 
&\rightarrow&
 \left( \frac{M_\Delta^*}{M_\Delta}\right)
\tilde{C}_{5,ij}  
\end{eqnarray}
where $i=V,A,VA$, and $j=1,2$.
The rest of the parameters of the SE formula are not modified.

\section{Tables of coefficients of the semi-empirical MEC formulas}

\label{apendice1}

In this appendix we provide the values of the coefficients,
$\tilde{C}_i(q)$, of the semi empirical MEC responses for $q=200$
MeV/c up to $q=2000$ MeC/c in steps of $\Delta q= 100$ MeV/c. These
tables can be interpolated to compute them for other
$q$-values. Alternatively, polynomial parametrizations of these
coefficient are provided in Appendix \ref{apendice2}.
The tables are included as additional material in plain text.

In table \ref{t0} we give the 
averaged width, $\Gamma(q)$, and averaged shift, $\Sigma(q)$,
 of the averaged $\Delta$ propagator, $G_{\rm av}(q,\omega)$,
for each one of the response functions of the kind $VV$, $AA$ or $VA$.

The coefficients of the SE MEC  are provided in tables 
\ref{t1} (for the $R_T^{VV}$ response function),
\ref{t2} (for $R_T^{AA}$),
\ref{t3} (for $R_{CC}^{VV}$),
\ref{t4} (for $R_{CC}^{AA}$),
\ref{t5} (for $R_{CL}^{AA}$),
\ref{t6} (for  $R_{LL}^{AA}$), and 
\ref{t7} (for $R_{T'}^{VA}$).

\begingroup
\squeezetable
\begin{table}[h]
\caption{Values of the $\Sigma$ and $\Gamma$ parameters of the averaged
  $\Delta$ propagator in the vector, axial and vector-axial
  responses.}
\label{t0}
\begin{tabular*}{\textwidth}{@{\extracolsep{\fill}}rdddddd}
\hline\hline\\[-2.5mm]
\multicolumn{1}{c}{q} & 
\multicolumn{1}{c}{$\Sigma_V$} & 
\multicolumn{1}{c}{$\Gamma_V$ } & 
\multicolumn{1}{c}{$\Sigma_A$} & 
\multicolumn{1}{c}{$\Gamma_A$} & 
\multicolumn{1}{c}{$\Sigma_{VA}$} & 
\multicolumn{1}{c}{$\Gamma_{VA} $}  \\ \hline
  200 &   429.56 &   411.88 &   370.98 &   354.59 &  -414.99 &   392.89 \\
  300 &   78.986 &   185.09 &   56.717 &   213.63 &   68.141 &   202.39 \\
  400 &   91.125 &   166.59 &   77.657 &   105.54 &   86.064 &   114.85 \\
  500 &   98.153 &   110.12 &   79.080 &   108.77 &   89.183 &   109.51 \\
  600 &   102.60 &   101.32 &   81.094 &   112.29 &   92.511 &   104.93 \\
  700 &   107.11 &   95.220 &   83.883 &   112.49 &   96.193 &   100.76 \\
  800 &   111.70 &   89.771 &   87.198 &   111.84 &   100.14 &   96.588 \\
  900 &   116.35 &   84.322 &   90.914 &   111.16 &   104.23 &   92.175 \\
  1000 &   120.93 &   79.042 &   94.968 &   110.81 &   108.29 &   87.599 \\
  1100 &   125.48 &   74.275 &   99.373 &   110.79 &   112.16 &   82.932 \\
  1200 &   130.43 &   70.570 &   104.18 &   110.83 &   115.85 &   78.412 \\
  1300 &   136.28 &   67.973 &   109.36 &   110.74 &   119.34 &   74.053 \\
  1400 &   143.48 &   66.060 &   114.90 &   110.19 &   122.58 &   69.760 \\
  1500 &   152.32 &   64.801 &   120.78 &   108.83 &   125.39 &   65.752 \\
  1600 &   163.36 &   63.297 &   126.84 &   106.84 &   127.73 &   61.963 \\
  1700 &   177.02 &   60.783 &   132.90 &   103.96 &   129.37 &   58.403 \\
  1800 &   193.85 &   56.980 &   138.53 &   100.53 &   130.07 &   55.603 \\
  1900 &   214.45 &   51.489 &   143.10 &   97.028 &   129.72 &   53.580 \\
  2000 &   239.30 &   44.367 &   145.85 &   94.244 &   128.13 &   52.692 \\ \hline\hline

\end{tabular*}
\end{table}
\endgroup

\begingroup
\squeezetable
\begin{table}[h]
\caption{Coefficients $\tilde{C}_{i}(q)$ 
of the $R_T^{VV}$ response function in the
  semi-empirical MEC formulas.}
\label{t1}
\begin{tabular*}{\textwidth}{@{\extracolsep{\fill}}rddddddddd}
\hline\hline\\[-2.5mm]
\multicolumn{1}{c}{q} & 
\multicolumn{1}{c}{$\tilde{C}_{1,V1}$ } & 
\multicolumn{1}{c}{$\tilde{C}_{1,V2}$ } & 
\multicolumn{1}{c}{$\tilde{C}_{2,V}$ } & 
\multicolumn{1}{c}{$\tilde{C}_{3,V}$ } & 
\multicolumn{1}{c}{$\tilde{C}_{4,V1}$ } & 
\multicolumn{1}{c}{$\tilde{C}_{4,V2}$ } & 
\multicolumn{1}{c}{$\tilde{C}_{5,V1}$ } & 
\multicolumn{1}{c}{$\tilde{C}_{5,V2}$ } & 
\multicolumn{1}{c}{$\tilde{C}_{6,V}$ } 
\\ \hline
  200 &  -5.6455 &   4.7254 &   324.95 &   0.8821 &   19.976 &  -11.856 &  -0.0750 &  -0.1005 &   4.9442 \\
  300 &  -2.3128 &   7.0572 &   277.05 &   1.6642 &   6.7713 &  -8.4869 &  -0.2942 &  -0.1056 &   8.2559 \\
  400 &   2.8522 &   5.6335 &   237.33 &   2.3559 &  -3.6706 &  -2.9271 &  -0.4448 &  -0.1328 &   10.339 \\
  500 &   4.6067 &   5.6932 &   207.99 &   2.9384 &  -4.6034 &  -2.4378 &  -0.4102 &  -0.2456 &   11.498 \\
  600 &   6.3810 &   7.3485 &   187.15 &   3.4589 &  -5.3142 &  -2.0497 &  -0.4561 &  -0.3564 &   12.148 \\
  700 &   8.6406 &   10.267 &   172.73 &   3.9693 &  -6.3515 &  -1.3441 &  -0.5438 &  -0.4778 &   12.560 \\
  800 &   11.229 &   14.080 &   163.24 &   4.5053 &  -7.4147 &  -0.2435 &  -0.6380 &  -0.5912 &   12.867 \\
  900 &   14.017 &   18.385 &   157.49 &   5.0833 &  -8.3548 &   1.2379 &  -0.7307 &  -0.6812 &   13.126 \\
  1000 &   16.967 &   22.966 &   154.81 &   5.7084 &  -9.1218 &   3.0420 &  -0.8307 &  -0.7465 &   13.364 \\
  1100 &   20.125 &   27.880 &   154.72 &   6.3784 &  -9.6915 &   5.1327 &  -0.9510 &  -0.7926 &   13.588 \\
  1200 &   23.578 &   33.383 &   156.96 &   7.0876 &  -10.030 &   7.5290 &  -1.1024 &  -0.8230 &   13.803 \\
  1300 &   27.448 &   39.763 &   161.41 &   7.8296 &  -10.104 &   10.278 &  -1.2937 &  -0.8428 &   14.007 \\
  1400 &   31.840 &   47.170 &   168.06 &   8.5967 &  -9.8987 &   13.412 &  -1.5289 &  -0.8607 &   14.200 \\
  1500 &   36.836 &   55.518 &   177.07 &   9.3820 &  -9.4335 &   16.925 &  -1.8104 &  -0.8945 &   14.380 \\
  1600 &   42.528 &   64.336 &   188.67 &   10.179 &  -8.7568 &   20.743 &  -2.1399 &  -0.9698 &   14.545 \\
  1700 &   49.031 &   72.952 &   203.28 &   10.980 &  -7.9275 &   24.742 &  -2.5173 &  -1.1085 &   14.696 \\
  1800 &   56.482 &   80.929 &   221.50 &   11.781 &  -6.9963 &   28.795 &  -2.9423 &  -1.3179 &   14.831 \\
  1900 &   65.053 &   88.622 &   244.16 &   12.577 &  -5.9638 &   32.999 &  -3.4146 &  -1.5896 &   14.948 \\
  2000 &   74.960 &   98.238 &   272.46 &   13.365 &  -4.7513 &   38.182 &  -3.9421 &  -2.0225 &   15.050 \\ \hline \hline
\end{tabular*}
\end{table}
\endgroup

\begingroup
\squeezetable
\begin{table}
\caption{Coefficients $\tilde{C}_{i}(q)$ 
of the $R_T^{AA}$ response function in the
  semi-empirical MEC formulas.}
\label{t2}
\begin{tabular*}{\textwidth}{@{\extracolsep{\fill}}rddddddddd}
\hline\hline\\[-2.5mm]
\multicolumn{1}{c}{q} & 
\multicolumn{1}{c}{$\tilde{C}_{1,A1}$ } & 
\multicolumn{1}{c}{$\tilde{C}_{1,A2}$ } & 
\multicolumn{1}{c}{$\tilde{C}_{2,A}$ } & 
\multicolumn{1}{c}{$\tilde{C}_{3,A}$ } & 
\multicolumn{1}{c}{$\tilde{C}_{4,A1}$ } & 
\multicolumn{1}{c}{$\tilde{C}_{4,A2}$ } & 
\multicolumn{1}{c}{$\tilde{C}_{5,A1}$ } & 
\multicolumn{1}{c}{$\tilde{C}_{5,A2}$ } & 
\multicolumn{1}{c}{$\tilde{C}_{6,A}$ } 
\\ \hline
  200 &   16.329 &   34.206 &   19.944 &   20.615 &   4.1488 &  -2.5516 &  -2.7292 &   2.6273 &   17.649 \\
  300 &   21.430 &   18.820 &   18.724 &   18.470 &   3.7251 &  -0.2755 &  -2.3497 &  -0.0523 &  16.196 \\
  400 &   22.912 &   23.973 &   16.754 &   15.620 &   2.7107 &   1.3642 &  -2.1103 &  -1.1070 &   13.495 \\
  500 &   24.292 &   28.051 &   14.811 &   13.089 &   2.4231 &   1.8619 &  -2.2025 &  -1.5055 &   10.924 \\
  600 &   25.597 &   31.450 &   13.062 &   11.054 &   2.2471 &   2.2170 &  -2.3253 &  -1.7242 &   8.8130 \\
  700 &   26.669 &   34.336 &   11.520 &   9.4730 &   2.0649 &   2.5229 &  -2.3886 &  -1.8179 &   7.1491 \\
  800 &   27.590 &   36.844 &   10.172 &   8.2615 &   1.8603 &   2.7740 &  -2.3935 &  -1.8096 &   5.8390 \\
  900 &   28.405 &   39.083 &   8.9866 &   7.3289 &   1.6293 &   2.9601 &  -2.3566 &  -1.7396 &   4.7773 \\
  1000 &   29.015 &   41.136 &   7.9396 &   6.6055 &   1.3711 &   3.0905 &  -2.2927 &  -1.6460 &   3.8973 \\
  1100 &   29.845 &   43.044 &   7.0164 &   6.0317 &   1.0980 &   3.1824 &  -2.2107 &  -1.5499 &   3.1401 \\
  1200 &   30.530 &   44.799 &   6.1995 &   5.5604 &   0.8124 &   3.2402 &  -2.1127 &  -1.4694 &   2.4634 \\
  1300 &   31.210 &   46.400 &   5.4813 &   5.1521 &   0.5207 &   3.2635 &  -1.9980 &  -1.4097 &   1.8359 \\
  1400 &   31.879 &   47.821 &   4.8477 &   4.7766 &   0.2225 &   3.2622 &  -1.8677 &  -1.3610 &   1.2352 \\
  1500 &   32.511 &   49.056 &   4.2924 &   4.4065 &  -0.0743 &   3.2381 &  -1.7222 &  -1.3142 &   0.6426 \\
  1600 &   33.070 &   50.100 &   3.8069 &   4.0198 &  -0.3670 &   3.1950 &  -1.5655 &  -1.2580 &   0.0461 \\
  1700 &   33.514 &   50.979 &   3.3820 &   3.5983 &  -0.6496 &   3.1351 &  -1.4033 &  -1.1864 &  -0.5625 \\
  1800 &   33.803 &   51.770 &   3.0103 &   3.1267 &  -0.9176 &   3.0642 &  -1.2428 &  -1.1018 &  -1.1922 \\
  1900 &   33.905 &   52.613 &   2.6875 &   2.5931 &  -1.1712 &   2.9866 &  -1.0965 &  -0.9982 &  -1.8438 \\
  2000 &   33.778 &   53.762 &   2.4071 &   1.9874 &  -1.3992 &   2.9084 &  -0.9692 &  -0.8808 &  -2.5212 \\ \hline \hline
\end{tabular*}
\end{table}
\endgroup

\begingroup
\squeezetable
\begin{table}
\caption{Coefficients $\tilde{C}_{i}(q)$ 
of the $R_{CC}^{VV}$ response function in the
  semi-empirical MEC formulas.}
\label{t3}
\begin{tabular*}{\textwidth}{@{\extracolsep{\fill}}rddddddddd}
\hline\hline\\[-2.5mm]
\multicolumn{1}{c}{q} & 
\multicolumn{1}{c}{$\tilde{C}_{1,V1}$ } & 
\multicolumn{1}{c}{$\tilde{C}_{1,V2}$ } & 
\multicolumn{1}{c}{$\tilde{C}_{2,V}$ } & 
\multicolumn{1}{c}{$\tilde{C}_{3,V}$ } & 
\multicolumn{1}{c}{$\tilde{C}_{4,V1}$ } & 
\multicolumn{1}{c}{$\tilde{C}_{4,V2}$ } & 
\multicolumn{1}{c}{$\tilde{C}_{5,V1}$ } & 
\multicolumn{1}{c}{$\tilde{C}_{5,V2}$ } & 
\multicolumn{1}{c}{$\tilde{C}_{6,V}$ } 
\\ \hline
   200 &   -0.1024 &    0.0854 &   8.5525 &   0.0476 &   0.3211 &  -0.1617 &   0.0030 &  -0.0135 &   0.0357 \\
  300 &   -0.0486 &    0.1498 &   6.7440 &   0.1319 &   0.1593 &  -0.1606 &  -0.0159 &  -0.0211 &   0.1003 \\
  400 &    0.0678 &    0.1505 &   5.4143 &   0.2642 &  -0.0385 &  -0.0810 &  -0.0463 &  -0.0248 &   0.1893 \\
  500 &    0.1578 &    0.1882 &   4.5372 &   0.4452 &  -0.0496 &  -0.0868 &  -0.0653 &  -0.0422 &   0.3005 \\
  600 &    0.3027 &    0.2904 &   3.9966 &   0.6710 &   0.0294 &  -0.0948 &  -0.1020 &  -0.0694 &   0.4303 \\
  700 &    0.5658 &    0.4717 &   3.6898 &   0.9354 &   0.2210 &  -0.0787 &  -0.1666 &  -0.1106 &   0.5732 \\
  800 &    1.0273 &    0.7286 &   3.5402 &   1.2320 &   0.5934 &  -0.0121 &  -0.2661 &  -0.1706 &   0.7238 \\
  900 &    1.8174 &    1.0250 &   3.4920 &   1.5559 &   1.2182 &   0.1411 &  -0.4121 &  -0.2582 &   0.8770 \\
  1000 &   3.1690 &   1.2899 &   3.5088 &   1.9032 &   2.1686 &   0.4230 &  -0.6208 &  -0.3881 &   1.0291 \\
  1100 &   5.5066 &   1.4027 &   3.5654 &   2.2711 &   3.5213 &   0.8816 &  -0.9112 &  -0.5782 &   1.1773 \\
  1200 &   9.5630 &   1.1584 &   3.6453 &   2.6577 &   5.3523 &   1.5678 &  -1.3018 &  -0.8488 &   1.3193 \\
  1300 &   16.468 &   0.2226 &   3.7376 &   3.0612 &   7.7257 &   2.5367 &  -1.8108 &  -1.2243 &   1.4536 \\
  1400 &   27.710 &  -1.8853 &   3.8351 &   3.4808 &   10.692 &   3.8489 &  -2.4530 &  -1.7339 &   1.5790 \\
  1500 &   44.980 &  -5.7623 &   3.9328 &   3.9156 &   14.285 &   5.5617 &  -3.2416 &  -2.4152 &   1.6947 \\
  1600 &   70.073 &  -12.111 &   4.0275 &   4.3654 &   18.052 &   7.7114 &  -4.1882 &  -3.3012 &   1.8002 \\
  1700 &   104.98 &  -21.738 &   4.1169 &   4.8302 &   23.426 &   10.286 &  -5.3006 &  -4.4046 &   1.8959 \\
  1800 &   152.13 &  -35.588 &   4.1997 &   5.3101 &   29.007 &   13.208 &  -6.5868 &  -5.7040 &   1.9812 \\
  1900 &   214.67 &  -54.766 &   4.2747 &   5.8056 &   35.314 &   16.426 &  -8.0633 &  -7.1889 &   2.0576 \\
  2000 &   296.65 &  -80.395 &   4.3415 &   6.3170 &   42.457 &   20.217 &  -9.7724 &  -9.1248 &   2.1249 \\ \hline \hline
\end{tabular*}
\end{table}
\endgroup

\begingroup
\squeezetable
\begin{table}
\caption{Coefficients $\tilde{C}_{i}(q)$ 
of the $R_{CC}^{AA}$ response function in the
  semi-empirical MEC formulas.}
\label{t4}
\begin{tabular*}{\textwidth}{@{\extracolsep{\fill}}rdddddddddd}
\hline\hline\\[-2.5mm]
\multicolumn{1}{c}{q} & 
\multicolumn{1}{c}{$\tilde{C}_{1,A1}$ } & 
\multicolumn{1}{c}{$\tilde{C}_{1,A2}$ } & 
\multicolumn{1}{c}{$\tilde{C}_{2,A1}$ } & 
\multicolumn{1}{c}{$\tilde{C}_{2,A2}$ } & 
\multicolumn{1}{c}{$\tilde{C}_{3,A}$ } & 
\multicolumn{1}{c}{$\tilde{C}_{4,A1}$ } & 
\multicolumn{1}{c}{$\tilde{C}_{4,A2}$ } & 
\multicolumn{1}{c}{$\tilde{C}_{5,A1}$ } & 
\multicolumn{1}{c}{$\tilde{C}_{5,A2}$ } & 
\multicolumn{1}{c}{$\tilde{C}_{6,A}$ } 
\\ \hline
  200 &   0.4310 &   0.6758 &   238.27 &   9.3999 &   0.3545 &   1.0876 &  -50.597 &   0.0504 &  -0.0292 &   0.3475 \\
  300 &   0.7425 &   0.6100 &   215.77 &   8.4094 &   0.3314 &  -8.2958 &  -11.926 &   0.0265 &   0.0090 &   0.0575 \\
  400 &   1.1073 &   1.3165 &   175.09 &   7.7175 &   0.3006 &  -7.1932 &  -8.0118 &   0.0064 &   0.0216 &  -0.1498 \\
  500 &   2.1427 &   2.1218 &   138.76 &   6.9936 &   0.2787 &  -4.9007 &  -7.7936 &   0.0210 &   0.0146 &  -0.3151 \\
  600 &   3.1362 &   3.1395 &   109.68 &   6.2763 &   0.2646 &  -0.4025 &  -8.1195 &   0.0515 &   0.0109 &  -0.4585 \\
  700 &   4.3281 &   4.3521 &   87.033 &   5.6002 &   0.2544 &   4.6146 &  -8.0345 &   0.0965 &   0.0144 &  -0.5908 \\
  800 &   5.7860 &   5.7550 &   69.523 &   4.9837 &   0.2442 &   10.156 &  -7.5209 &   0.1605 &   0.0296 &  -0.7057 \\
  900 &   7.6019 &   7.3393 &   55.929 &   4.4317 &   0.2305 &   16.577 &  -6.5162 &   0.2447 &   0.0588 &  -0.8128 \\
  1000 &   9.9183 &   9.1025 &   45.328 &   3.9429 &   0.2106 &   24.215 &  -4.9410 &   0.3495 &   0.1041 &  -0.9201 \\
  1100 &   12.958 &   11.016 &   37.025 &   3.5122 &   0.1822 &   33.432 &  -2.6731 &   0.4694 &   0.1641 &  -1.0171 \\
  1200 &   17.041 &   13.007 &   30.456 &   3.1346 &   0.1430 &   44.503 &   0.4040 &   0.5954 &   0.2368 &  -1.1184 \\
  1300 &   22.616 &   14.950 &   25.230 &   2.8037 &   0.0918 &   57.725 &   4.4288 &   0.7159 &   0.3149 &  -1.2640 \\
  1400 &   30.266 &   16.646 &   21.041 &   2.5137 &   0.0271 &   73.368 &   9.5238 &   0.8128 &   0.3868 &  -1.3986 \\
  1500 &   40.702 &   17.843 &   17.664 &   2.2593 &  -0.0514 &   91.621 &   15.780 &   0.8644 &   0.4368 &  -1.5547 \\
  1600 &   54.774 &   18.212 &   14.919 &   2.0359 &  -0.1444 &   112.59 &   23.235 &   0.8495 &   0.4422 &  -1.7426 \\
  1700 &   73.477 &   17.381 &   12.677 &   1.8392 &  -0.2522 &   136.52 &   31.950 &   0.7379 &   0.3723 &  -1.9275 \\
  1800 &   97.967 &   14.957 &   10.833 &   1.6658 &  -0.3739 &   163.50 &   41.932 &   0.5062 &   0.1975 &  -2.1423 \\
  1900 &   129.63 &   10.517 &   9.3095 &   1.5124 &  -0.5101 &   193.73 &   53.264 &   0.1143 &  -0.1323 &  -2.3821 \\
  2000 &   170.12 &   3.6516 &   8.0440 &   1.3765 &  -0.6595 &   227.56 &   66.185 &  -0.4731 &  -0.6773 &  -2.6522 \\ \hline \hline
\end{tabular*}
\end{table}
\endgroup

\begingroup
\squeezetable
\begin{table}
\caption{Coefficients $\tilde{C}_{i}(q)$ 
of the  $R_{CL}^{AA}$ response function in the
  semi-empirical MEC formulas.}
\label{t5}
\begin{tabular*}{\textwidth}{@{\extracolsep{\fill}}rdddddddddd}
\hline\hline\\[-2.5mm]
\multicolumn{1}{c}{q} & 
\multicolumn{1}{c}{$\tilde{C}_{1,A1}$ } & 
\multicolumn{1}{c}{$\tilde{C}_{1,A2}$ } & 
\multicolumn{1}{c}{$\tilde{C}_{2,A1}$ } & 
\multicolumn{1}{c}{$\tilde{C}_{2,A2}$ } & 
\multicolumn{1}{c}{$\tilde{C}_{3,A}$ } & 
\multicolumn{1}{c}{$\tilde{C}_{4,A1}$ } & 
\multicolumn{1}{c}{$\tilde{C}_{4,A2}$ } & 
\multicolumn{1}{c}{$\tilde{C}_{5,A1}$ } & 
\multicolumn{1}{c}{$\tilde{C}_{5,A2}$ } & 
\multicolumn{1}{c}{$\tilde{C}_{6,A}$ } 
\\ \hline
  200 &  -1.1765 &  -2.5379 &  -72.678 &  -7.2538 &   0.0806 &   5.5173 &   34.373 &   0.0563 &   0.1300 &  -0.7486 \\
  300 &  -2.1851 &  -2.0364 &  -91.567 &  -7.3504 &  -0.1306 &   9.9664 &   13.202 &   0.1902 &  -0.0182 &  -0.6646 \\
  400 &  -3.0251 &  -3.6397 &  -86.641 &  -7.0749 &  -0.2579 &   8.1204 &   10.129 &   0.1989 &   0.0688 &  -0.4324 \\
  500 &  -4.9236 &  -5.0983 &  -75.322 &  -6.5790 &  -0.3193 &   6.9727 &   9.2048 &   0.1870 &   0.1269 &  -0.2418 \\
  600 &  -6.3387 &  -6.6909 &  -63.463 &  -5.9990 &  -0.3393 &   2.5570 &   9.3724 &   0.1727 &   0.1655 &  -0.1346 \\
  700 &  -7.8449 &  -8.3950 &  -52.859 &  -5.4096 &  -0.3340 &  -2.7198 &   9.1389 &   0.1527 &   0.1909 &  -0.1047 \\
  800 &  -9.5543 &  -10.225 &  -43.896 &  -4.8494 &  -0.3134 &  -8.5236 &   8.4471 &   0.1251 &   0.2045 &  -0.1183 \\
  900 &  -11.573 &  -12.190 &  -36.470 &  -4.3350 &  -0.2830 &  -15.121 &   7.2693 &   0.0907 &   0.2064 &  -0.1513 \\
  1000 &  -14.046 &  -14.312 &  -30.392 &  -3.8718 &  -0.2457 &  -22.835 &   5.5497 &   0.0531 &   0.2008 &  -0.1864 \\
  1100 &  -17.186 &  -16.579 &  -25.435 &  -3.4593 &  -0.2019 &  -32.018 &   3.1768 &   0.0182 &   0.1931 &  -0.2067 \\
  1200 &  -21.292 &  -18.932 &  -21.379 &  -3.0945 &  -0.1506 &  -42.966 &   0.0363 &  -0.0051 &   0.1917 &  -0.1904 \\
  1300 &  -26.782 &  -21.257 &  -18.053 &  -2.7730 &  -0.0910 &  -55.972 &  -4.0071 &  -0.0072 &   0.2058 &  -0.1377 \\
  1400 &  -34.193 &  -23.366 &  -15.321 &  -2.4898 &  -0.0212 &  -71.313 &  -9.0779 &   0.0268 &   0.2493 &  -0.0681 \\
  1500 &  -44.183 &  -25.025 &  -13.066 &  -2.2405 &   0.0598 &  -89.177 &  -15.266 &   0.1140 &   0.3370 &   0.0141 \\
  1600 &  -57.540 &  -25.924 &  -11.196 &  -2.0209 &   0.1528 &  -109.72 &  -22.625 &   0.2718 &   0.4888 &   0.1558 \\
  1700 &  -75.203 &  -25.715 &  -9.6367 &  -1.8273 &   0.2584 &  -133.11 &  -31.192 &   0.5271 &   0.7308 &   0.3566 \\
  1800 &  -98.277 &  -24.042 &  -8.3357 &  -1.6560 &   0.3769 &  -159.48 &  -40.991 &   0.9038 &   1.0907 &   0.5645 \\
  1900 &  -128.11 &  -20.522 &  -7.2402 &  -1.5045 &   0.5089 &  -189.04 &  -52.100 &   1.4318 &   1.6084 &   0.8363 \\
  2000 &  -166.30 &  -14.797 &  -6.3199 &  -1.3699 &   0.6537 &  -222.20 &  -64.787 &   2.1540 &   2.3542 &   1.1126 \\ \hline \hline
\end{tabular*}
\end{table}
\endgroup

\begingroup
\squeezetable
\begin{table}
\caption{Coefficients $\tilde{C}_{i}(q)$ 
of the  $R_{LL}^{AA}$ response function in the
  semi-empirical MEC formulas.}
\label{t6}
\begin{tabular*}{\textwidth}{@{\extracolsep{\fill}}rdddddddddd}
\hline\hline\\[-2.5mm]
\multicolumn{1}{c}{q} & 
\multicolumn{1}{c}{$\tilde{C}_{1,A1}$ } & 
\multicolumn{1}{c}{$\tilde{C}_{1,A2}$ } & 
\multicolumn{1}{c}{$\tilde{C}_{2,A1}$ } & 
\multicolumn{1}{c}{$\tilde{C}_{2,A2}$ } & 
\multicolumn{1}{c}{$\tilde{C}_{3,A}$ } & 
\multicolumn{1}{c}{$\tilde{C}_{4,A1}$ } & 
\multicolumn{1}{c}{$\tilde{C}_{4,A2}$ } & 
\multicolumn{1}{c}{$\tilde{C}_{5,A1}$ } & 
\multicolumn{1}{c}{$\tilde{C}_{5,A2}$ } & 
\multicolumn{1}{c}{$\tilde{C}_{6,A}$ } 
\\ \hline
  200 &   7.7088 &   17.531 &   233.20 &   5.0374 &   9.4317 &   3.1687 &  -49.648 &  -1.2839 &   1.2848 &   3.4218 \\
  300 &   10.129 &   9.2250 &   180.29 &   6.0908 &   8.1410 &  -5.1516 &  -17.177 &  -1.1784 &   0.1189 &   2.1615 \\
  400 &   10.742 &   11.672 &   130.90 &   6.3740 &   6.7068 &  -5.1196 &  -11.029 &  -0.9983 &  -0.4267 &   1.5098 \\
  500 &   12.296 &   13.556 &   95.868 &   6.1594 &   5.5049 &  -5.8216 &  -8.7191 &  -0.9352 &  -0.6065 &   1.2160 \\
  600 &   13.590 &   15.271 &   71.787 &   5.7313 &   4.5619 &  -2.0130 &  -8.5029 &  -0.8949 &  -0.6911 &   1.1226 \\
  700 &   14.933 &   16.968 &   55.051 &   5.2304 &   3.8319 &   3.0303 &  -8.0879 &  -0.8569 &  -0.7349 &   1.1446 \\
  800 &   16.477 &   18.749 &   43.149 &   4.7251 &   3.2680 &   8.6441 &  -7.2951 &  -0.8243 &  -0.7630 &   1.2170 \\
  900 &   18.343 &   20.667 &   34.439 &   4.2463 &   2.8272 &   15.023 &  -6.0701 &  -0.8030 &  -0.7775 &   1.3022 \\
  1000 &   20.673 &   22.776 &   27.928 &   3.8069 &   2.4761 &   22.503 &  -4.3435 &  -0.8019 &  -0.7903 &   1.3936 \\
  1100 &   23.665 &   25.080 &   22.940 &   3.4113 &   2.1877 &   31.380 &  -2.0133 &  -0.8286 &  -0.8101 &   1.4126 \\
  1200 &   27.589 &   27.524 &   19.060 &   3.0580 &   1.9416 &   42.013 &   1.0546 &  -0.8948 &  -0.8527 &   1.4160 \\
  1300 &   32.827 &   29.998 &   15.983 &   2.7448 &   1.7231 &   54.657 &   4.9885 &  -1.0084 &  -0.9300 &   1.3692 \\
  1400 &   39.872 &   32.313 &   13.511 &   2.4679 &   1.5207 &   69.579 &   9.9124 &  -1.1826 &  -1.0605 &   1.2548 \\
  1500 &   49.320 &   34.251 &   11.500 &   2.2234 &   1.3272 &   86.942 &   15.909 &  -1.4322 &  -1.2638 &   1.0643 \\
  1600 &   61.902 &   35.509 &   9.8547 &   2.0074 &   1.1366 &   106.95 &   23.046 &  -1.7716 &  -1.5598 &   0.8203 \\
  1700 &   78.500 &   35.763 &   8.4949 &   1.8163 &   0.9445 &   129.79 &   31.376 &  -2.2227 &  -1.9720 &   0.5642 \\
  1800 &   100.17 &   34.688 &   7.3611 &   1.6472 &   0.7488 &   155.55 &   40.907 &  -2.8139 &  -2.5330 &   0.2057 \\
  1900 &   128.23 &   31.939 &   6.4108 &   1.4973 &   0.5476 &   184.45 &   51.726 &  -3.5732 &  -3.2833 &  -0.2214 \\
  2000 &   164.22 &   27.217 &   5.6116 &   1.3640 &   0.3392 &   216.84 &   64.070 &  -4.5204 &  -4.2713 &  -0.6653 \\ \hline \hline
\end{tabular*}
\end{table}
\endgroup

\begingroup
\squeezetable
\begin{table}
\caption{Coefficients $\tilde{C}_{i}(q)$ 
of the  $R_{T'}^{VA}$ response function in the
  semi-empirical MEC formulas.}
\label{t7}
\begin{tabular*}{\textwidth}{@{\extracolsep{\fill}}rdddddddddd}
\hline\hline\\[-2.5mm]
\multicolumn{1}{c}{q} & 
\multicolumn{1}{c}{$\tilde{C}_{1,VA1}$ } & 
\multicolumn{1}{c}{$\tilde{C}_{1,VA2}$ } & 
\multicolumn{1}{c}{$\tilde{C}_{2,VA}$ } & 
\multicolumn{1}{c}{$\tilde{C}_{3,VA}$ } & 
\multicolumn{1}{c}{$\tilde{C}_{4,VA1}$ } & 
\multicolumn{1}{c}{$\tilde{C}_{4,VA2}$ } & 
\multicolumn{1}{c}{$\tilde{C}_{5,VA1}$ } & 
\multicolumn{1}{c}{$\tilde{C}_{5,VA2}$ } & 
\multicolumn{1}{c}{$\tilde{C}_{6,VA}$ } & 
\multicolumn{1}{c}{$\tilde{C}_{6,AV}$ } 
\\ \hline
   200 &   3.0775 &   7.0669 &   33.839 &   4.1248 &  -122.58 &   175.36 &  -0.28107 &   0.16341 &  -84.995 &   135.39 \\
  300 &   6.0856 &   7.0257 &   39.591 &   5.2873 &  -45.017 &   68.720 &  -0.36476 &  -0.00935 &  -5.9848 &   24.604 \\
  400 &   8.2977 &   9.3332 &   40.713 &   5.6595 &   10.642 &  -13.723 &  -0.41794 &  -0.17344 &   6.0440 &   6.4537\\
  500 &   10.413 &   12.083 &   39.821 &   5.6071 &   20.192 &  -31.555 &  -0.44973 &  -0.30008 &   5.8446 &   5.3355\\
  600 &   12.682 &   15.166 &   38.151 &   5.3665 &   17.331 &  -30.514 &  -0.49020 &  -0.39654 &   4.1826 &   6.3500\\
  700 &   14.967 &   18.484 &   36.245 &   5.0677 &   12.947 &  -26.221 &  -0.52323 &  -0.46538 &   2.9376 &   6.8790\\
  800 &   17.214 &   21.943 &   34.390 &   4.7752 &   9.2326 &  -22.152 &  -0.53893 &  -0.50361 &   2.1534 &   6.8773\\
  900 &   19.389 &   25.464 &   32.660 &   4.5129 &   6.4570 &  -19.001 &  -0.53699 &  -0.51340 &   1.6290 &   6.6367\\
  1000 &   21.465 &   29.016 &   31.110 &   4.2871 &   4.4528 &  -16.072 &  -0.51836 &  -0.49662 &   1.2863 &   6.2528\\
  1100 &   23.421 &   32.597 &   29.762 &   4.0941 &   3.0130 &  -15.132 &  -0.48558 &  -0.45727 &   1.0445 &   5.8248\\
  1200 &   25.233 &   36.210 &   28.602 &   3.9272 &   1.9657 &  -14.031 &  -0.44114 &  -0.39951 &   0.86212 &   5.3968\\
  1300 &   26.891 &   39.863 &   27.639 &   3.7794 &   1.1886 &  -13.267 &  -0.38738 &  -0.32521 &   0.71212 &   4.9943\\
  1400 &   28.388 &   43.548 &   26.853 &   3.6443 &   0.59663 &  -12.723 &  -0.32579 &  -0.23678 &   0.59011 &   4.6091\\
  1500 &   29.714 &   47.274 &   26.240 &   3.5167 &   0.13074 &  -12.320 &  -0.25785 &  -0.13292 &   0.48717 &   4.2444\\
  1600 &   30.863 &   51.023 &   25.787 &   3.3926 &  -0.24908 &  -12.001 &  -0.18497 &  -0.01558 &   0.39962 &   3.8983\\
  1700 &   31.828 &   54.804 &   25.479 &   3.2691 &  -0.57026 &  -11.726 &  -0.10735 &   0.11481 &   0.32699 &   3.5639\\
  1800 &   32.598 &   58.687 &   25.307 &   3.1443 &  -0.85053 &  -11.474 &  -0.02521 &   0.25836 &   0.26413 &   3.2461\\
  1900 &   33.166 &   62.755 &   25.264 &   3.0170 &  -1.1045 &  -11.225 &   0.06088 &   0.41597 &   0.21063 &   2.9409\\
  2000 &   33.523 &   67.137 &   25.330 &   2.8868 &  -1.3401 &  -10.971 &   0.15142 &   0.59231 &   0.16352 &   2.6510\\\hline \hline
\end{tabular*}
\end{table}
\endgroup

\clearpage

\section{Parametrizations of nuclear response coefficients }
\label{apendice2}

In this appendix we provide parametrizations of the $q$-dependence of the 
semi-empirical formula coefficients, $\tilde{C}_i(q)$,
and of the averaged width, $\Gamma(q)$, and averaged  shift, $\Sigma(q)$,
 of the averaged $\Delta$ propagator, $G_{\rm av}(q,\omega)$,
for each one of the response functions 
in the semi-empirical MEC formulas.
We provide two kind of parametrizations. Most of the coefficients
 allow a polynomial 
parametrization:
\begin{equation} \label{polinomio1}
\tilde{C}_i(q)
 = \sum_{j=0}^8 a_j \left( \frac{q}{2m_N}\right)^j
\end{equation}
where the maximum range of the polynomial is
eight for some of the coefficients, 
but in many cases the polynomial degree is less or much lesser than eight. 
The factors in the polynomial are dimensionless.

For some of the coefficients the parametrization is better written as a
polynomial of the inverse variable $2m_N/q$
\begin{equation}  \label{polinomio2}
\tilde{C}_i(q)
 = \sum_{j=0}^8 a_j \left( \frac{2m_N}{q}\right)^j
\end{equation}
These are labeled with a star symbol (*), in the tables below.

The parameters of the 
averaged width, $\Gamma(q)$, and averaged  shift, $\Sigma(q)$,
are given in Tab. \ref{RNME0}.

Parameters are provided in tables 
\ref{RNME1} (for the $R_T^{VV}$ response function),
\ref{RNME2} (for $R_T^{AA}$),
\ref{RNME3} (for $R_{CC}^{VV}$),
\ref{RNME4} (for $R_{CC}^{AA}$),
\ref{RNME5} (for $R_{CL}^{AA}$),
\ref{RNME6} (for  $R_{LL}^{AA}$), and 
\ref{RNME7} (for $R_{T'}^{VA}$).

\begingroup
\squeezetable
\begin{table}[h]
\caption{Parameters of the effective shift, $\Sigma$, and width,
  $\Gamma$, of the averaged $\Delta$ propagator in the vector, axial
  and vector-axial responses in the polynomial interpolations given by Eq.
  (\ref{polinomio1}), or Eq. (\ref{polinomio2}) for the ones labeled
  with a star symbol $*$.  }
\label{RNME0}
\begin{tabular*}{\textwidth}{@{\extracolsep{\fill}}ldddddddd}
\hline\hline
\multicolumn{1}{c}{} & 
\multicolumn{1}{c}{$a_{0}$} & 
\multicolumn{1}{c}{$a_{1}$} & 
\multicolumn{1}{c}{$a_{2}$} & 
\multicolumn{1}{c}{$a_{3}$} & 
\multicolumn{1}{c}{$a_{4}$} & 
\multicolumn{1}{c}{$a_{5}$} & 
\multicolumn{1}{c}{$a_{6}$} \\ \hline
$\Sigma_{V}$ & 60.5191 & 174.869 & -142.624 & 0.0 & 64.8675 & 51.5966 & 0.0 \\
$\Gamma_{V}$ & 1450.93 & -14248.6 & 60480.2 & -132124 & 155882 & -94328.1 & 22938.6 \\ \hline
$\Sigma_{A}$* & 93.7987 & -207.022 & 880.16 & -1497.82 & 1345.81 & -472.81 & 0.0 \\
$\Gamma_{A}$ & 48.5674 & 457.116 & -983.684 & 0.0 & 2539.27 & -3068.47 & 1104.8 \\ \hline
$\Sigma_{VA}$* & 83.6648 & -43.5277 & 349.139 & -500.553 & 371.11 & -129.884 & 0.0 \\
$\Gamma_{VA}$* & 173.605 & -566.125 & 2174.47 & -4916.66 & 5945.42 & -3691.91 & 935.097 \\ \hline
\end{tabular*}
\end{table}
\endgroup

\begingroup
\squeezetable
\begin{table}
\caption{Parameters of the coefficients $\tilde{C}_{i}(q)$ 
of the response function $R_{T}^{VV}$ in the interpolations
  given by Eq.  (\ref{polinomio1}), or Eq. (\ref{polinomio2}) for the
  ones labeled with a star symbol $*$.  }
\label{RNME1}
\begin{tabular*}{\textwidth}{@{\extracolsep{\fill}}l@{\kern 3mm}d@{\kern 4mm}d@{\kern 4mm}d@{\kern 3mm}d@{\kern 3mm}d@{\kern 1mm}d@{\kern 2mm}d@{\kern 0mm}d@{\hspace{-1mm}}d@{\kern -1mm}}
\hline\hline
\multicolumn{1}{c}{} & 
\multicolumn{1}{c}{$a_{0}$} & 
\multicolumn{1}{c}{$a_{1}$} & 
\multicolumn{1}{c}{$a_{2}$} & 
\multicolumn{1}{c}{$a_{3}$} & 
\multicolumn{1}{c}{$a_{4}$} & 
\multicolumn{1}{c}{$a_{5}$} &
\multicolumn{1}{c}{$a_{6}$} &
\multicolumn{1}{c}{$a_{7}$} & 
\multicolumn{1}{c}{$a_{8}$} \\ \hline \\[-2mm]
$\tilde{C}_{1,V1}$ & -19.6664 & 174.729  & -504.898 & 878.162  & -695.892 & 230.403  & 0.0      & 0.0      & 0.0     \\
$\tilde{C}_{1,V2}$ & 0.0      & 253.264  & -2441.02 & 9243.9   & -14669.3 & 5825.93  & 10974.9  & -13547.1 & 4446.27 \\
$\tilde{C}_{2,V}$  & 464.634  & -1606.42 & 3069.48  & -2681.38 & 991.981  & 0.0      & 0.0      & 0.0      & 0.0     \\
$\tilde{C}_{3,V}$  & 0.0      & 10.3393  & 0.0      & 2.04851  & 0.0      & 0.0      & 0.0      & 0.0      & 0.0     \\
$\tilde{C}_{4,V1}$ & 94.6881  & -1119.14 & 5033.78  & -11612.8 & 14309.6  & -8943.98 & 2231.48  & 0.0      & 0.0     \\
$\tilde{C}_{4,V2}$& -32.638  & 281.151  & -1007.61 & 1735.12  & -1334.5  & 390.521  & 0.0      & 0.0      & 0.0     \\
$\tilde{C}_{5,V1}$ & 0.0      & -2.28139 & 4.09607  & -5.10766 & 0.0      & 0.0      & 0.0      & 0.0      & 0.0     \\
$\tilde{C}_{5,V2}$ & 0.0      & 0.0      & -4.75457 & 0.0      & 12.9405  & -9.69383 & 0.0      & 0.0      & 0.0     \\
$\tilde{C}_{6,V}$ & -8.6321  & 188.024  & -702.668 & 1404.74  & -1543.33 & 882.04   & -205.234 & 0.0      & 0.0     \\ \hline \hline
\end{tabular*}
\end{table}
\endgroup

\begingroup
\squeezetable
\begin{table}
\caption{Parameters of the coefficients $\tilde{C}_{i}(q)$ 
of the response function $R_{T}^{AA}$ in the polynomial interpolations 
  given by Eq.  (\ref{polinomio1}) or Eq. (\ref{polinomio2}) for the
  ones labeled with a star symbol $*$.  }
\label{RNME2}
\begin{tabular*}{\textwidth}{@{\extracolsep{\fill}}ldddddddd@{\kern -4mm}}
\hline\hline
\multicolumn{1}{c}{} & 
\multicolumn{1}{c}{$a_{0}$} & 
\multicolumn{1}{c}{$a_{1}$} & 
\multicolumn{1}{c}{$a_{2}$} & 
\multicolumn{1}{c}{$a_{3}$} & 
\multicolumn{1}{c}{$a_{4}$} & 
\multicolumn{1}{c}{$a_{5}$} &
\multicolumn{1}{c}{$a_{6}$} \\ \hline \\[-2mm]
$\tilde{C}_{1,A1}$* & 40.1435 & -6.64283 & 0.0 & 0.354216 & -0.0596866 & 0.00286741 & 0.0 \\
$\tilde{C}_{1,A2}$* & 70.1364 & -18.7631 & 0.0 & 1.64276 & -0.448886 & 0.0478086 & -0.0017799 \\
$\tilde{C}_{2,A}$* & -4.81568 & 8.57371 & -1.05139 & 0.0446577 & 0.0 & 0.0 & 0.0 \\
$\tilde{C}_{3,A}$* & -1.70277 & 4.96453 & -0.275884 & 0.0 & 0.0 & 0.0 & 0.0 \\
$\tilde{C}_{4,A1}$* & -7.1971 & 8.61063 & -2.84088 & 0.399384 & -0.0192562 & 0.0 & 0.0 \\
$\tilde{C}_{4,A2}$* & 2.41265 & 1.1986 & -0.572642 & 0.0682446 & -0.00286093 & 0.0 & 0.0 \\
$\tilde{C}_{5,A1}$* & 3.60383 & -7.63457 & 3.56018 & -0.757029 & 0.0746929 & -0.00277361 & 0.0 \\
$\tilde{C}_{5,A2}$* & 0.564127 & -2.1691 & 0.636593 & -0.0652588 & 0.00261604 & 0.0 & 0.0 \\
$\tilde{C}_{6,A}$* & -6.78958 & 6.19498 & -0.386024 & 0.0 & 0.0 & 0.0 & 0.0 \\ \hline \hline
\end{tabular*}
\end{table}
\endgroup

\begingroup
\squeezetable
\begin{table}
\caption{Parameters of the coefficients $\tilde{C}_{i}(q)$ 
of the response function $R_{CC}^{VV}$ in the polynomial interpolations
  given by Eq.  (\ref{polinomio1}), or Eq. (\ref{polinomio2}) for the
  ones labeled with a star symbol $*$.  }
\label{RNME3}
\begin{tabular*}{\textwidth}{@{\extracolsep{\fill}}l@{\kern -1mm}dd@{\kern 2mm}d@{\kern 5mm}d@{\kern 3mm}d@{\kern 1mm}d@{\kern 0 mm}d@{\kern 0mm}d@{\hspace{-1mm}}d@{\kern -2mm}}
\hline\hline
\multicolumn{1}{c}{} & 
\multicolumn{1}{c}{$a_{0}$} & 
\multicolumn{1}{c}{$a_{1}$} & 
\multicolumn{1}{c}{$a_{2}$} & 
\multicolumn{1}{c}{$a_{3}$} & 
\multicolumn{1}{c}{$a_{4}$} & 
\multicolumn{1}{c}{$a_{5}$} &
\multicolumn{1}{c}{$a_{6}$} &
\multicolumn{1}{c}{$a_{7}$} & 
\multicolumn{1}{c}{$a_{8}$} \\ \hline \\[-2mm]
$\tilde{C}_{1,V1}$  & 5.86212 & -159.987 & 1683.15 & -9120.25 & 28225.1 & -51256.7 & 53538.9 & -29301.5 & 6584.11 \\ 
$\tilde{C}_{1,V2}$  & 0.0 & 10.2676 & -141.292 & 839.978 & -2670.46 & 4936.0 & -5039.63 & 2519.75 & -504.43 \\
$\tilde{C}_{2,V}$ & 13.7234 & -61.873 & 132.704 & -119.937 & 39.5931 & 0.0 & 0.0 & 0.0 & 0.0 \\
$\tilde{C}_{3,V}$  & -0.0686547 & 0.0 & 8.18253 & -2.41841 & 0.0 & 0.0 & 0.0 & 0.0 & 0.0 \\
$\tilde{C}_{4,V1}$  & 1.75335 & -23.1728 & 127.716 & -386.742 & 628.412 & -421.798 & 107.647 & 0.0 & 0.0 \\
$\tilde{C}_{4,V2}$ & 0.0 & -4.25877 & 36.7999 & -123.11 & 166.971 & -60.689 & 0.0 & 0.0 & 0.0 \\
$\tilde{C}_{5,V1}$  & 0.0 & -1.50012 & 8.60699 & -14.8142 & 0.0 & 0.0 & 0.0 & 0.0 & 0.0 \\
$\tilde{C}_{5,V2}$  & 0.0 & 0.542587 & -6.74484 & 22.0466 & -29.4463 & 6.74147 & 0.0 & 0.0 & 0.0 \\
$\tilde{C}_{6,V}$ & 0.0 & -0.288279 & 5.91647 & 0.0 & -10.6428 & 9.91934 & -2.86346 & 0.0 & 0.0 \\ \hline \hline
\end{tabular*}
\end{table}
\endgroup

\begingroup
\squeezetable
\begin{table}
\caption{Parameters of the coefficients $\tilde{C}_{i}(q)$ 
of the response function $R_{CC}^{AA}$ in the polynomial interpolations
  given by Eq.  (\ref{polinomio1}), or Eq. (\ref{polinomio2}) for the
  ones labeled with a star symbol $*$.  }
\label{RNME4}
\begin{tabular*}{\textwidth}{@{\extracolsep{\fill}}ld@{\kern 3mm}d@{\kern 3mm}d@{\kern 3mm}d@{\kern 3mm}d@{\kern 3mm}d@{\kern 3mm}d@{\kern -2mm}}
\hline\hline
\multicolumn{1}{c}{} & 
\multicolumn{1}{c}{$a_{0}$} & 
\multicolumn{1}{c}{$a_{1}$} & 
\multicolumn{1}{c}{$a_{2}$} & 
\multicolumn{1}{c}{$a_{3}$} & 
\multicolumn{1}{c}{$a_{4}$} & 
\multicolumn{1}{c}{$a_{5}$} &
\multicolumn{1}{c}{$a_{6}$} \\ \hline \\[-2mm]
$\tilde{C}_{1,A1}$ & 1.00967 & -13.0344 & 73.0221 & 0.0 & -198.704 & 259.352 & 0.0 \\ 
$\tilde{C}_{1,A2}$ & 3.83926 & -62.2338 & 400.672 & -1071.58 & 1612.63 & -1160.11 & 288.513 \\
$\tilde{C}_{2,A1}$* & -39.489 & 44.8644 & 1.70487 & -0.355919 & 0.0 & 0.0 & 0.0 \\
$\tilde{C}_{2,A2}$* & -2.17605 & 4.26239 & -0.589857 & 0.0284662 & 0.0 & 0.0 & 0.0 \\
$\tilde{C}_{3,A}$ & 0.491785 & -1.6485 & 4.50652 & -5.29804 & 1.47136 & 0.0 & 0.0 \\
$\tilde{C}_{4,A1}$ & 0.0 & -39.1279 & 69.8619 & 155.687 & 0.0 & 0.0 & 0.0 \\
$\tilde{C}_{4,A2}$ & -237.745 & 3006.35 & -14949.7 & 36613.0 & -47206.4 & 30846.5 & -8019.86 \\
$\tilde{C}_{5,A1}$ & 0.0888813 & -0.283259 & -1.87097 & 7.84059 & 0.0 & -5.55097 & 0.0 \\
$\tilde{C}_{5,A2}$ & -0.408277 & 5.44882 & -22.9732 & 37.91 & -20.0388 & 0.0 & 0.0 \\
$\tilde{C}_{6,A}$ & 1.04713 & -8.02409 & 14.0568 & -12.9651 & 3.55436 & 0.0 & 0.0 \\ \hline \hline
\end{tabular*}
\end{table}
\endgroup

\begingroup
\squeezetable
\begin{table}
\caption{Parameters of the coefficients $\tilde{C}_{i}(q)$ 
of the  response function $R_{CL}^{AA}$ in the polynomial interpolations
  given by Eq.  (\ref{polinomio1}), or Eq. (\ref{polinomio2}) for the
  ones labeled with a star symbol $*$.  }
\label{RNME5}
\begin{tabular*}{\textwidth}{@{\extracolsep{\fill}}ld@{\kern 3mm}d@{\kern 3mm}d@{\kern 3mm}d@{\kern 3mm}d@{\kern 3mm}d@{\kern 3mm}d@{\kern -1mm}}
\hline\hline
\multicolumn{1}{c}{} & 
\multicolumn{1}{c}{$a_{0}$} & 
\multicolumn{1}{c}{$a_{1}$} & 
\multicolumn{1}{c}{$a_{2}$} & 
\multicolumn{1}{c}{$a_{3}$} & 
\multicolumn{1}{c}{$a_{4}$} & 
\multicolumn{1}{c}{$a_{5}$} &
\multicolumn{1}{c}{$a_{6}$} \\ \hline \\[-2mm]
$\tilde{C}_{1,A1}$ & 0.0 & -5.53339 & -53.5517 & 0.0 & 177.554 & -239.114 & 0.0 \\ 
$\tilde{C}_{1,A2}$ & -9.61627 & 139.949 & -899.062 & 2402.54 & -3417.33 & 2375.71 & -613.648 \\
$\tilde{C}_{2,A1}$* & 0.0 & 11.7941 & -25.2288 & 6.74752 & -0.71334 & 0.0274023 & 0.0 \\
$\tilde{C}_{2,A2}$* & 2.15609 & -4.27799 & 0.628945 & -0.0298366 & 0.0 & 0.0 & 0.0 \\
$\tilde{C}_{3,A}$ & 0.636771 & -7.06185 & 17.1626 & -16.5615 & 6.28956 & 0.0 & 0.0 \\
$\tilde{C}_{4,A1}$ & 0.0 & 56.6646 & -111.747 & -126.814 & 0.0 & 0.0 & 0.0 \\
$\tilde{C}_{4,A2}$ & 136.9 & -1630.92 & 7986.45 & -19326.8 & 24651.6 & -16013.6 & 4146.06 \\
$\tilde{C}_{5,A1}$ & 0.0 & 1.19682 & -1.25693 & -2.63388 & 0.0 & 3.99421 & 0.0 \\
$\tilde{C}_{5,A2}$ & 0.325959 & -3.9354 & 18.934 & -32.2136 & 18.3759 & 0.0 & 0.0 \\
$\tilde{C}_{6,A}$ & -1.6945 & 10.4321 & -24.2097 & 21.6826 & -5.45423 & 0.0 & 0.0 \\ \hline \hline
\end{tabular*}
\end{table}
\endgroup

\begingroup
\squeezetable
\begin{table}
\caption{Parameters of the coefficients $\tilde{C}_{i}(q)$ 
of the response function $R_{LL}^{AA}$ in the polynomial interpolations
  given by Eq.  (\ref{polinomio1}), or Eq. (\ref{polinomio2}) for the
  ones labeled with a star symbol $*$.  }
\label{RNME6}
\begin{tabular*}{\textwidth}{@{\extracolsep{\fill}}l@{\kern 1mm}dd@{\kern 2mm}d@{\kern 2mm}d@{\kern 2mm}d@{\kern -1mm}d@{\kern -1mm}d@{\kern 0mm}d@{\hspace{-1mm}}d@{\kern -4mm}}
\hline\hline
\multicolumn{1}{c}{} & 
\multicolumn{1}{c}{$a_{0}$} & 
\multicolumn{1}{c}{$a_{1}$} & 
\multicolumn{1}{c}{$a_{2}$} & 
\multicolumn{1}{c}{$a_{3}$} & 
\multicolumn{1}{c}{$a_{4}$} & 
\multicolumn{1}{c}{$a_{5}$} &
\multicolumn{1}{c}{$a_{6}$} &
\multicolumn{1}{c}{$a_{7}$} & 
\multicolumn{1}{c}{$a_{8}$} 
\\ \hline \\[-2mm]
$\tilde{C}_{1,A1}$ & 6.14051 & 16.8939 & 26.5549 & 0.0 & -131.934 & 203.693 & 0.0 & 0.0 & 0.0 \\ 
$\tilde{C}_{1,A2}$& 141.573 & -2569.7 & 19929.8 & -81032.4 & 193118 & -278723 & 239909 & -113349 & 22609.3 \\
$\tilde{C}_{2,A1}$* & -12.4591 & 12.5106 & 6.0512 & -0.490028 & 0.0 & 0.0 & 0.0 & 0.0 & 0.0 \\
$\tilde{C}_{2,A2}$* & -2.2448 & 4.431 & -0.710191 & 0.0341804 & 0.0 & 0.0 & 0.0 & 0.0 & 0.0 \\
$\tilde{C}_{3,A}$ & 13.7017 & -45.4469 & 66.7382 & -45.3372 & 10.9484 & 0.0 & 0.0 & 0.0 & 0.0 \\
$\tilde{C}_{4,A1}$ & 0.0 & -34.6322 & 55.3018 & 156.919 & 0.0 & 0.0 & 0.0 & 0.0 & 0.0 \\
$\tilde{C}_{4,A2}$ & -204.783 & 2442.11 & -11760.5 & 28204.8 & -35811.5 & 23149.8 & -5969.69 & 0.0 & 0.0 \\
$\tilde{C}_{5,A1}$ & -1.63891 & 4.02606 & -6.87094 & 5.50556 & 0.0 & -4.39721 & 0.0 & 0.0 & 0.0 \\
$\tilde{C}_{5,A2}$ & 3.52402 & -29.492 & 67.8688 & -60.0326 & 14.9714 & 0.0 & 0.0 & 0.0 & 0.0 \\
$\tilde{C}_{6,A}$ & 6.74744 & -43.2203 & 113.243 & -117.526 & 40.5894 & 0.0 & 0.0 & 0.0 & 0.0 \\ \hline \hline
\end{tabular*}
\end{table}
\endgroup

\clearpage

\begingroup
\squeezetable
\begin{table}[h]
\caption{Parameters of the coefficients $\tilde{C}_{i}(q)$ 
of the response function $R_{T'}^{VA}$ in the polynomial interpolations
  given by Eq.  (\ref{polinomio1}), or Eq. (\ref{polinomio2}) for the
  ones labeled with a star symbol $*$.  }
\label{RNME7}
\begin{center}
\begin{tabular*}{\textwidth}{@{\extracolsep{\fill}}l@{\extracolsep{\fill}\kern -2mm}
d@{\kern 1mm}
d@{\kern 1mm}
d@{\kern 1mm}
d@{\kern -1mm}
d@{\kern 0mm}
d@{\kern -3mm}
d@{\kern -3mm}
d@{\hspace{-2mm}}
d@{\kern -4mm}}
\hline\hline
\multicolumn{1}{c}{} & 
\multicolumn{1}{c}{$a_{0}$} & 
\multicolumn{1}{c}{$a_{1}$} & 
\multicolumn{1}{c}{$a_{2}$} & 
\multicolumn{1}{c}{$a_{3}$} & 
\multicolumn{1}{c}{$a_{4}$} & 
\multicolumn{1}{c}{$a_{5}$} &
\multicolumn{1}{c}{$a_{6}$} &
\multicolumn{1}{c}{$a_{7}$} & 
\multicolumn{1}{c}{$a_{8}$} \\ \hline \\[-2mm]
$\tilde{C}_{1,VA1}$  & -12.5654 & 271.856 & -1772.34 & 7170.88 & -16850.5 & 23866.6 & -20160.5 & 9354.45 & -1834.88 \\ 
$\tilde{C}_{1,VA2}$  & 25.7693 & -401.628 & 3111.18 & -11803.4 & 27278.8 & -38962.8 & 33541.9 & -15937.6 & 3209.45 \\
$\tilde{C}_{2,VA}$  & 0.981663 & 516.73 & -2461.18 & 5506.71 & -6578.11 & 4042.59 & -1002.31 & 0.0 & 0.0 \\
$\tilde{C}_{3,VA}$  & -0.698901 & 70.138 & -272.492 & 456.781 & -355.974 & 105.235 & 0.0 & 0.0 & 0.0 \\
$\tilde{C}_{4,VA1}$ * & -12.6139 & 27.5451 & -27.9095 & 15.1272 & -3.53758 & 0.346934 & -0.0120447 & 0.0 & 0.0 \\
$\tilde{C}_{4,VA2}$ * & -21.627 & 21.438 & -11.7492 & 0.0 & 0.549239 & -0.0443672 & 0.0 & 0.0 & 0.0 \\
$\tilde{C}_{5,VA1}$  & -0.239117 & 0.0 & -6.79621 & 17.8376 & -15.5393 & 4.77589 & 0.0 & 0.0 & 0.0 \\
$\tilde{C}_{5,VA2}$  & 0.570313 & -3.93619 & 0.0 & 12.945 & -14.1847 & 4.9816 & 0.0 & 0.0 & 0.0 \\
$\tilde{C}_{6,VA}$ * & -3.54219 & 8.09376 & -6.67611 & 2.807 & -0.500736 & 0.035343 & -0.000846361 & 0.0 & 0.0 \\
$\tilde{C}_{6,AV}$ * & -0.47772 & 0.0 & 6.24771 & -3.31642 & 0.616557 & -0.0426815 & 0.000953283 & 0.0 & 0.0 \\ \hline \hline
\end{tabular*}
\end{center}
\end{table}
\endgroup



\end{document}